\documentclass[twocolumn]{aa}
\newcommand{\Mtor}{M_{\text{tor}}}
\newcommand{\Rtor}{R_{\text{tor}}}
\newcommand{\Mbh}{M_{\text{c}}}
\newcommand{\PGPU}{$\varphi-$GPU}

\usepackage{graphicx}
\usepackage{natbib}
\usepackage{txfonts}
\usepackage{hyperref}
\usepackage{cleveref}[2012/02/15]
\usepackage[normalem]{ulem}
\hypersetup{colorlinks = true, citecolor = {blue}, urlcolor= {blue}}
\usepackage{diagbox}
\usepackage{siunitx}
\usepackage{enumerate}
\usepackage{comment}
\makeatletter
\renewcommand*\aa@pageof{, page \thepage{} of \pageref*{LastPage}}
\makeatother

\usepackage{xcolor}

\begin{document}

\title{Global $m=1$ slow mode in near-Keplerian self-gravitating torus: \\ applications to stellar nuclear disks and AGN molecular tori
}

\author{Elena~Bannikova\inst{1,2}, Volodymyr Akhmetov\inst{3,4}, Peter Berczik \inst{4},  Serhii Skolota\inst{2,5}, \\ Massimo Capaccioli \thanks{Deceased 2026.} $^{\dagger}$\inst{1},  Maryna Ishchenko\inst{4,6}
}
\institute{INAF - Astronomical Observatory of Capodimonte, Salita Moiariello 16, I-80131, Naples, Italy
          \\ \email{\href{mailto:olena.bannikova@inaf.it}{olena.bannikova@inaf.it}}
    \and                
    Institute of Radio Astronomy, National Academy of Sciences of Ukraine, Mystetstv 4, UA-61002 Kharkiv, Ukraine
    \and 
    INAF--Osservatorio Astrofisico di Torino, Via Osservatorio 20, I-10025, Pino Torinese, Italy 
    \and                
    Main Astronomical Observatory, National Academy of Sciences of Ukraine, 27 Akademika Zabolotnoho St, UA-03143 Kyiv, Ukraine
    \and
    V.N. Karazin Kharkiv National University, Svobody Sq. 4, UA-61022, Kharkiv, Ukraine
    \and
    National University of Life and Environmental Sciences of Ukraine, Kyiv, Ukraine    }

\date{Received xxx / Accepted xxx}
 
\abstract    
{Global $m=1$ asymmetries are observed in a variety of self-gravitating astrophysical systems and are often interpreted as large-scale slow modes in near-Keplerian potentials. Prominent examples include eccentric nuclear disks in galactic centers, such as the double nucleus of M31. However, the dynamical origin and long-term stability of such modes remain poorly understood.}
{We investigate the dynamical evolution and stability of a self-gravitating, collisionless torus orbiting a dominant central mass, with the aim of determining whether a slow non-axisymmetric ($m=1$) mode can arise spontaneously.}
{We performed a suite of direct $N$-body simulations exploring a range of torus-to-central mass ratios and different initial particle distributions. The calculations were carried out with the high-order Hermite GPU integrator ($\varphi-$GPU), enabling us to follow the long-term dynamical evolution of systems with a large number of particles.}
{We find that a global slow $m=1$ mode forms spontaneously from initially axisymmetric configurations without imposed perturbations. The lopsided structure is sustained by coherent apsidal alignment of orbits and persists over secular timescales. Its maintenance requires hierarchical nonlinear coupling of low-order modes, including the participation of the $m=3$ component, as well as a sufficient vertical thickness of the torus, indicating that the instability is inherently three-dimensional. As a dynamical consequence of the long-lived overdensity, the central mass acquires an essential displacement with respect to the system barycenter.}
{Our results demonstrate that a long-lived global $m=1$ mode can arise naturally in a geometrically thick self-gravitating torus orbiting a central mass. The mechanism identified here provides a dynamical framework for understanding eccentric nuclear disks, such as the double nuclei of M31 and NGC~4486B, as well as the molecular tori in AGNs, and suggests that the resulting lopsided asymmetry may produce observable offsets of the central supermassive black hole.}


\titlerunning{Global $m=1$ slow mode in a near-Keplerian self-gravitating torus}
\authorrunning{E.~Bannikova et al.}
\maketitle

\section{Introduction}\label{sec:Intr}

Lopsided ($m=1$) modes are a common feature of stellar systems, appearing from galactic disks to the immediate vicinity of supermassive black holes (SMBHs). Such modes represent global eccentric distortions and can manifest as off-centered density enhancements, double nuclei, or one-armed spirals. 
They are found in different classes of astrophysical systems, including large-scale lopsided disks in spiral galaxies \citep{1995ApJ...447...82R, 2009PhR...471...75J} and eccentric nuclear stellar disks around SMBHs, observed as nuclei with asymmetric surface-brightness distributions, such as in M~31 \citep{1974ApJ...194..257L, 1993AJ....106.1436L, 1998AJ....116.2263L, 1999ApJ...522..772K}. 
A double (eccentric) nucleus is also present in the compact elliptical galaxy NGC~4486B \citep{1996ApJ...471L..79L, 2005ApJ...631..280B, Tahmasebzadeh_2025}. 
High-resolution HST observations have revealed that asymmetric or double nuclei are not rare among early-type galaxies, with estimated occurrence rates of order 10\% \citep{2005AJ....129.2138L}. Recent ALMA observations reveal that molecular tori in nearby AGN are not axisymmetric. High-resolution maps of NGC 613 show that the central depletion region is displaced from the AGN position, indicating an $m=1$ asymmetry in the torus \citep{2026A&A...705A.124C}. 
These findings motivate a detailed investigation of the formation and dynamical evolution of global $m=1$ modes.

A number of theoretical and numerical studies have shown that near-Keplerian stellar systems can support slowly precessing $m=1$ configurations. Early interest in this problem was largely motivated by the interpretation of the nuclear stellar disk in M~31. In particular, \citet{1995AJ....110..628T} proposed that the observed double nucleus of M~31 can be understood as an eccentric stellar disk associated with a global lopsided ($m=1$) mode. In this framework, the apparent double structure arises because stars spend more time near the apocenter than the pericenter, producing an asymmetric surface-brightness distribution with two apparent peaks. This interpretation naturally led to a number of theoretical models aimed at reproducing such lopsided configurations. 

The secular dynamics of slow m=1 configurations in thin near-Keplerian disks was investigated by \citet{Sridhar1999} and developed further through the linear normal-mode analysis of \citet{Tremaine_2001}. The latter showed that low-mass disks orbiting a dominant central object can support stable modes whose pattern frequencies are much smaller than the orbital frequency. Their persistence depends on the ability of self-gravity to maintain coherent apsidal precession across the disk. In particular, self-gravity-dominated discrete modes are most readily supported in narrow rings, whereas broad radial distributions are less favorable because differential precession reduces global apsidal coherence.

 In many models of eccentric nuclear disks, however, the lopsided geometry was imposed through the initial conditions or generated by an external perturbation.
 \citet{2001ApJ...555L..25J} found no linear $m=1$ instability in razor-thin annular disks orbiting a dominant central mass. Instead, finite-amplitude eccentric distortions were introduced, after which the system evolved into long-lived, slowly precessing lopsided states, with pattern speed increasing approximately linearly with the disk-to-central mass ratio.
\citet{2003ApJ...599..237P} constructed eccentric-disk models for M~31 by superposing Keplerian stellar orbits around a central mass and fitting a parameterized distribution function to reproduce the observed photometric and kinematic properties. The $m=1$ structure is prescribed through the eccentricity profile, with orbits computed in a fixed Keplerian potential without including disk self-gravity.
The problem was further explored using high-resolution $N$-body simulations by \citet{2001A&A...371..409B}, who modeled the nuclear disk of M~31 as a self-gravitating stellar system orbiting a SMBH. Their three-dimensional simulations showed that an initially eccentric, apsidally aligned disk can remain stable and long-lived.
Subsequent self-consistent orbit-superposition models were presented in \citep{2004ApJ...611..245S} and \citep{2013MNRAS.431...80B}, constructing equilibrium eccentric disks from aligned Keplerian orbits around a central mass, typically neglecting disk self-gravity. These models reproduce the observed structure but treat the $m=1$ configuration as prescribed. \citet{2007MNRAS.382..419S} investigated global 
$m=1$ modes in an isolated, purely stellar galactic disk. Their linear analysis considered a thin exponential disk, while the $N$-body simulations employed an initially axisymmetric Kuzmin–Toomre model. They showed that collective self-gravity can reduce differential precession and support slowly evolving lopsided modes, while the simulations demonstrated the spontaneous emergence of a global $m=1$ asymmetry. However, the formation and long-term evolution of such a mode in a near-Keplerian system dominated by a central mass were not addressed. Spontaneous dipole instabilities have also recently been identified in spherical self-gravitating systems, where their growth is associated with an inflection in the distribution function \citep{2025ApJ...993..150D}.

More recent studies have largely followed similar lines, considering eccentric nuclear disks in which the $m=1$ geometry is imposed through initial conditions or external perturbations. $N$-body simulations show that a coherent mode can be maintained when apsidal alignment is present initially, while orbit-based and secular approaches construct such configurations by design \citep{2021ApJ...920..149W,2021MNRAS.503.2713R}. External mechanisms, including SMBH recoil and galaxy interactions, can also produce eccentric disks through imposed perturbations \citep{2021ApJ...921L..12A,2024MNRAS.52711458R}. Recent studies indicate that self-gravity can help maintain coherent 
$m=1$ modes by counteracting differential precession, although long-lived global modes may require additional conditions \citep{2018ApJ...853..141M,Lithwick2026}.

Following our earlier results \citep{2012MNRAS.424..820B, 2017FrASS...4...60B, Bannikova2021,Bannikova2023}, we revisit this problem by focusing on the intrinsic dynamical role of torus self-gravity. In contrast to models in which the $m = 1$ geometry is imposed or externally triggered, we consider an initially axisymmetric toroidal distribution of massive particles orbiting a central mass. Using high-resolution direct 3D $N$-body simulations, we demonstrate that a global $m = 1$ mode emerges spontaneously after the system reaches a quasi-equilibrium state, without imposed asymmetry or external perturbations. We further investigate the conditions for the long-term persistence of this mode. Finally, we apply our results to stellar nuclear disks in M~31 and NGC~4486B, as well as to the molecular torus in NGC~613.

\section{Initial conditions and integration procedure}
\label{sec:ini}

\begin{figure}[h!]
\centering
\includegraphics[width = 44mm]{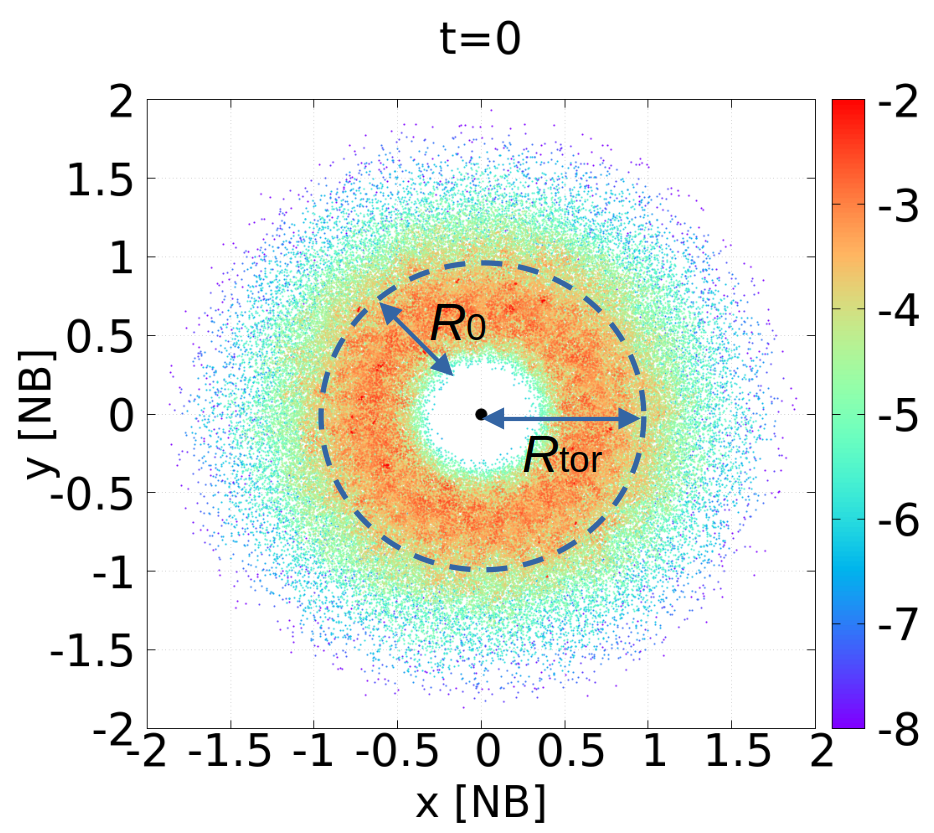}
\includegraphics[width = 44mm]{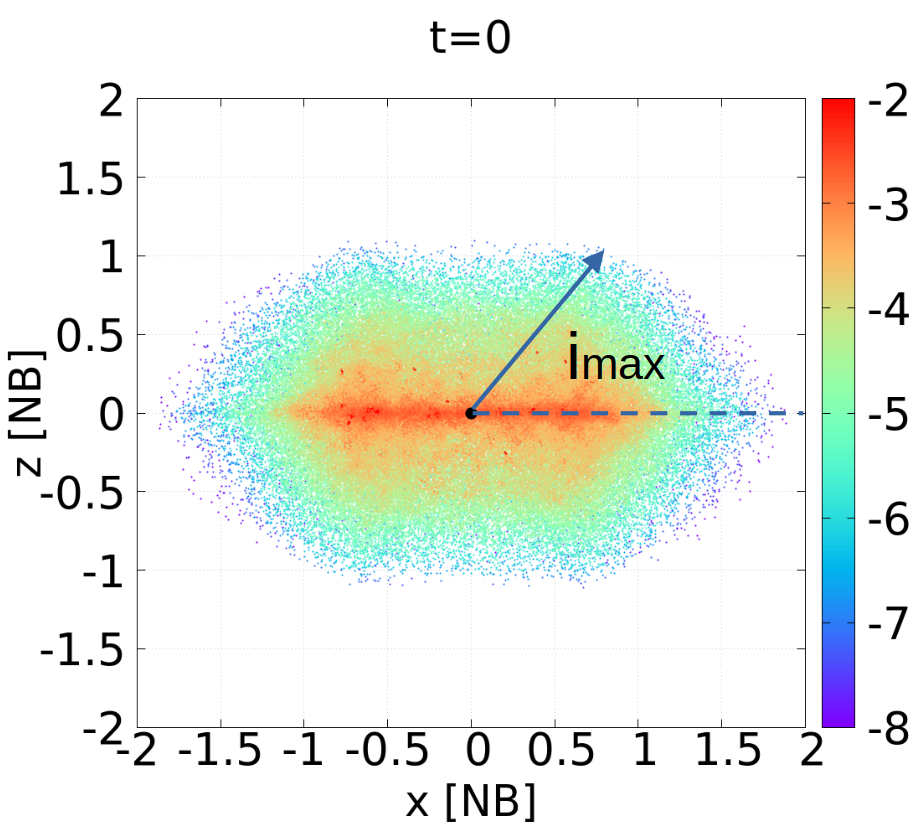}
\caption{An example of the initial density distribution. The arrows  show the key parameters of the system.}
\label{fig:density_00-i60}
\end{figure}

We consider a toroidal distribution of $N$ massive particles of equal mass $m$ orbiting a central mass ($\Mbh$).  We choose a random initial distribution of particles in Keplerian elements: the semi-major axes of all particles $a_k$ are randomly distributed within the interval \mbox{[$\Rtor - R_0$,.., $\Rtor +R_0$]}, with eccentricities in the range $e_k=[0,..,e_\text{max}]$. In this case, the mean radius of the particle distribution is located in the equatorial plane and corresponds to the major radius of the torus $\Rtor$. The characteristic radius $R_0$ represents the minor torus radius which determines the width of the particle spread (Fig.~\ref{fig:density_00-i60}). 
The toroidal structure is produced by a spread of orbital inclinations, resulting in a geometrically thick configuration. We assume that the inclinations of particle orbits are distributed in a range $i_k = [-i_\text{max},..,i_\text{max}]$. 
The longitudes of ascending nodes ($\Omega_k$) and true anomalies ($\nu_k$) are randomized within the interval $[0,..,2\pi]$. We fix the argument of periapsis $\omega_k=0$ for all runs. We also perform a control run with $\omega_k$ randomly distributed in the interval $[0, .., 2\pi]$, which shows that this does not affect the main results.

We adopt a system of units (NB) in which $G = 1$, \mbox{$M_c = 1$}, and $\Rtor = 1$. In these units, the mean orbital period of the torus, corresponding to a particle orbit in the equatorial plane with semi-major axis $a = \Rtor$, is $T_\text{orb} = 2\pi$. In the following, we use a dimensionless time $t$ normalized to the orbital period of the torus.
The torus mass $\Mtor$ is then expressed as a fraction of the central mass. For example, $\Mtor = 0.1$ means that the torus mass is 10\% of the central mass. This choice of torus mass in the canonical model is motivated by its relevance to the nuclear disk of M~31.

For the dynamical integration of the system, we employ a high-order parallel $N$-body code \PGPU\footnote{$N$-body code \PGPU: \\~\url{ https://github.com/berczik/phi-GPU-mole}} which is based on the Hermite integration scheme with hierarchical individual block time steps \citep{Berczik2011,BSW2013}. Our simulations may be sensitive to the integration time step, which depends on the parameter $\eta$ \citep{MA1992}. The integration parameter $\eta$ is chosen to ensure accurate conservation of the total energy, as verified in our previous simulations \citep{Bannikova2021}. We adopt here $\eta=0.007$, which provides a good compromise between computational cost and integration accuracy.

\begin{table}[htbp!]
\caption{The geometrical parameters of the system.}
\centering
\begin{tabular}{ccccccc}
\hline
Run & $e_\text{max}$ & $i_\text{max}$ [$^{\circ}$]  &  $R_0$ [NB]  \\
\hline
0.5-60-0.3 & 0.5 & 60 &  0.3  \\
0.5-45-0.3 & 0.5 & 45 &  0.3 \\
0.5-30-0.3 & 0.5 & 30 &  0.3 \\
0.2-60-0.3 & 0.2 & 60 &  0.3 \\
0.1-60-0.3 & 0.1 & 60 &  0.3 \\
0.36-60-0.3 & 0.3643 & 60 &  0.3 \\
0.8-60-0.3 & 0.8 & 60 &  0.3   \\
0.5-60-0.6 & 0.5 & 60 &  0.6 \\
 0.5-60-0.9 & 0.5 & 60 &  0.9 \\
\hline
\end{tabular}
\tablefoot{For all these runs the torus mass $\Mtor = 0.1$, the particle number $N=128k$, softening parameter \mbox{$\epsilon=10^{-2}$}.}
\label{tab:sec1:runs}
\end{table} 

We conducted a series of experiments, which can be divided into three categories. In the main set of simulations, we vary the geometric parameters of the initial distribution: $e_\text{max}$, $i_\text{max}$, and $R_0$. The names of the corresponding runs include these parameters, i.e. ‘run-$e_\text{max}$-$i_\text{max}$-$R_0$’. A list of these experiments is presented in Table~\ref{tab:sec1:runs}. In this set, the torus mass is fixed to $\Mtor = 0.1$. We included also the runs with initial $e_\text{max} = 0.1,0.3643$ that corresponds to the negative and zero values of initial virial (see Appendix~\ref{app:D}).
An additional set of runs is performed for fixed geometric parameters ($e_\text{max} = 0.5$, $i_\text{max} = 60^\circ$, and $R_0 = 0.3$), while varying the number of particles ($N = 64k, 256k$), the softening parameter ($\epsilon = 10^{-3}, 10^{-4}$), and the level of symmetry of the particle distribution. These runs are performed to verify that these parameters do not significantly affect the system dynamics; the corresponding results are presented in Appendix~\ref{App:A}. We also perform runs for different torus masses ($\Mtor = 0.06, 0.01$), the results of which are discussed in Sect.~\ref{sec:canonical}.

\section{Spontaneous formation of the \textit{m}=1 mode}
\label{sec:canonical}

In this section, we first show how an initially symmetric configuration spontaneously develops a large-scale asymmetry, and then examine its dynamical consequences and its dependence on the torus mass.

\subsection{Formation of the overdensity}
\label{subsec:form}

In the first subsection, we investigate the evolution of the density distribution of the torus for a mass $\Mtor=0.1$, corresponding to run-0.5-60-0.3 (the first line in Table~\ref{tab:sec1:runs}). To visualize the particle distribution, we construct density distribution maps following the procedure described in \citep{Bannikova2021}. 
\begin{figure}[h!]
\centering
\includegraphics[width = 44mm]{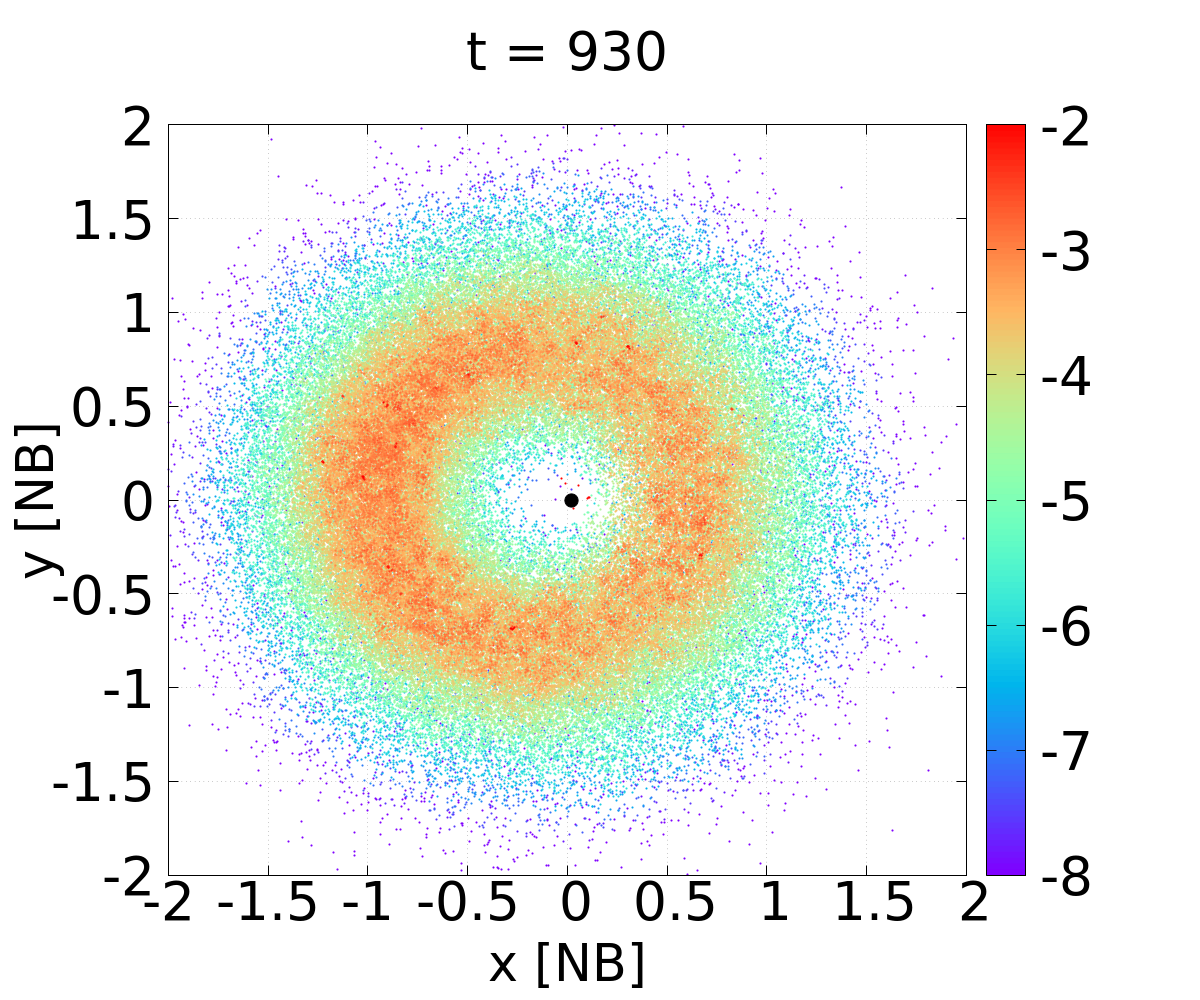}
\includegraphics[width = 44mm]{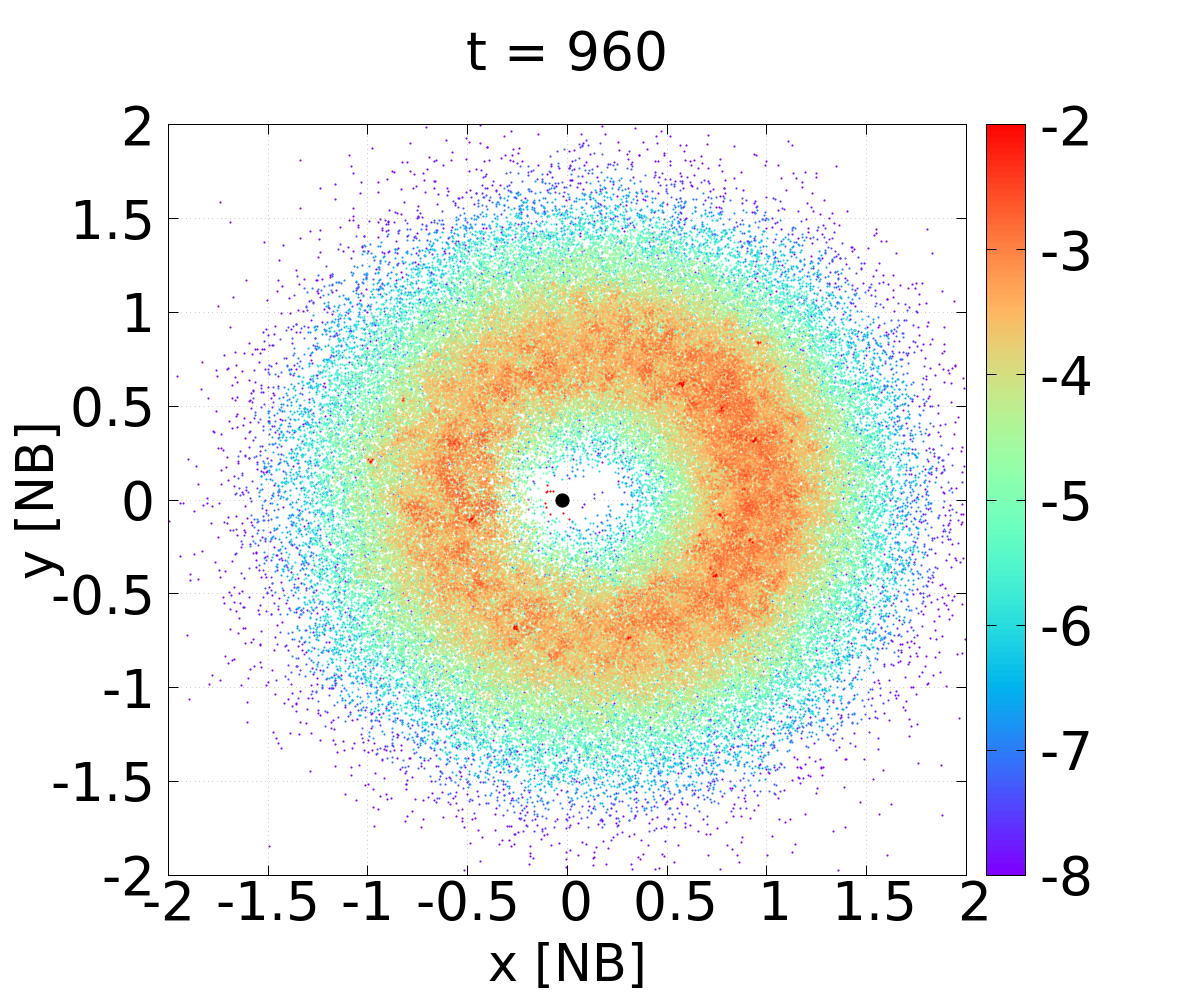}
\includegraphics[width = 44mm]{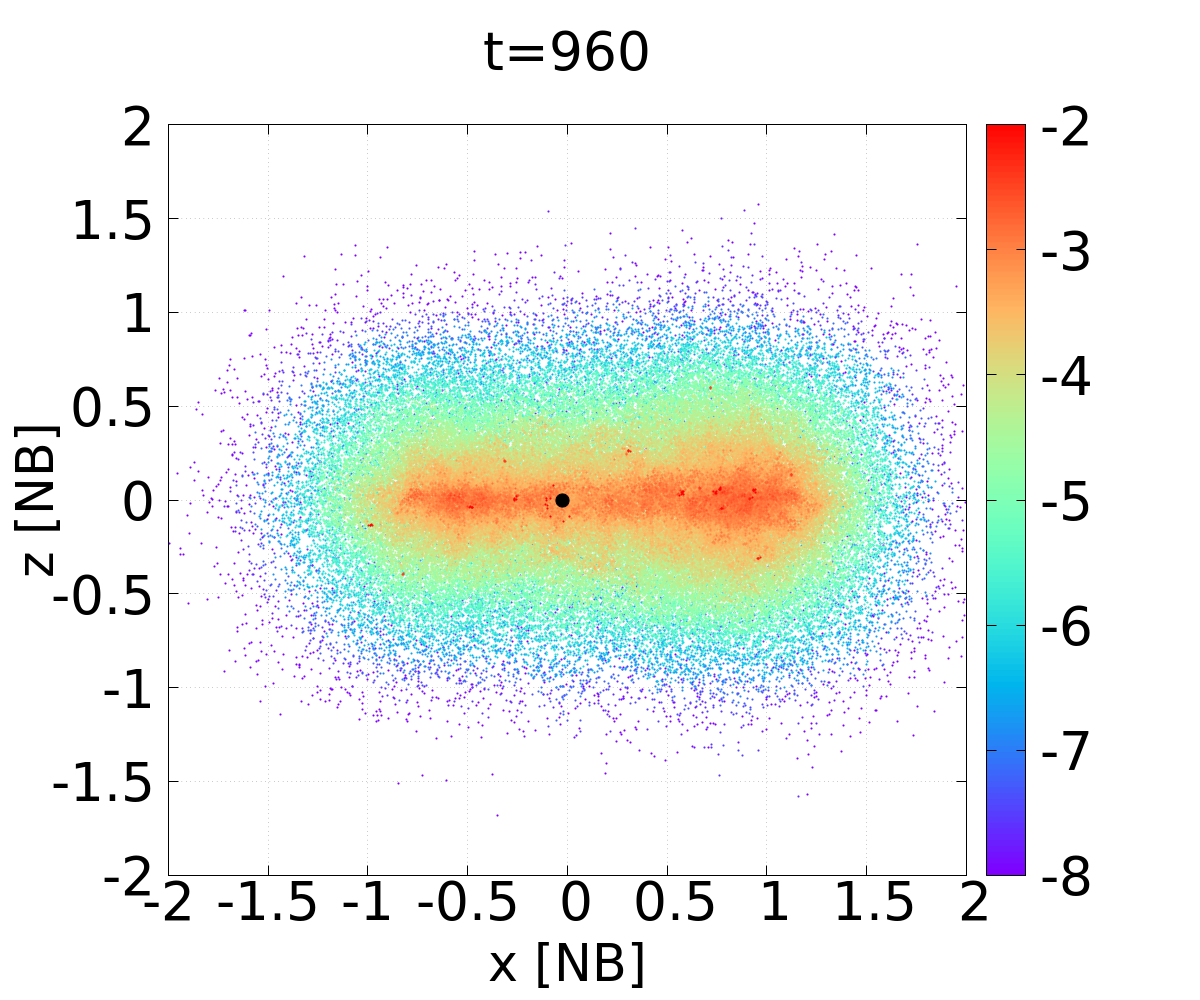}
\includegraphics[width = 44mm]{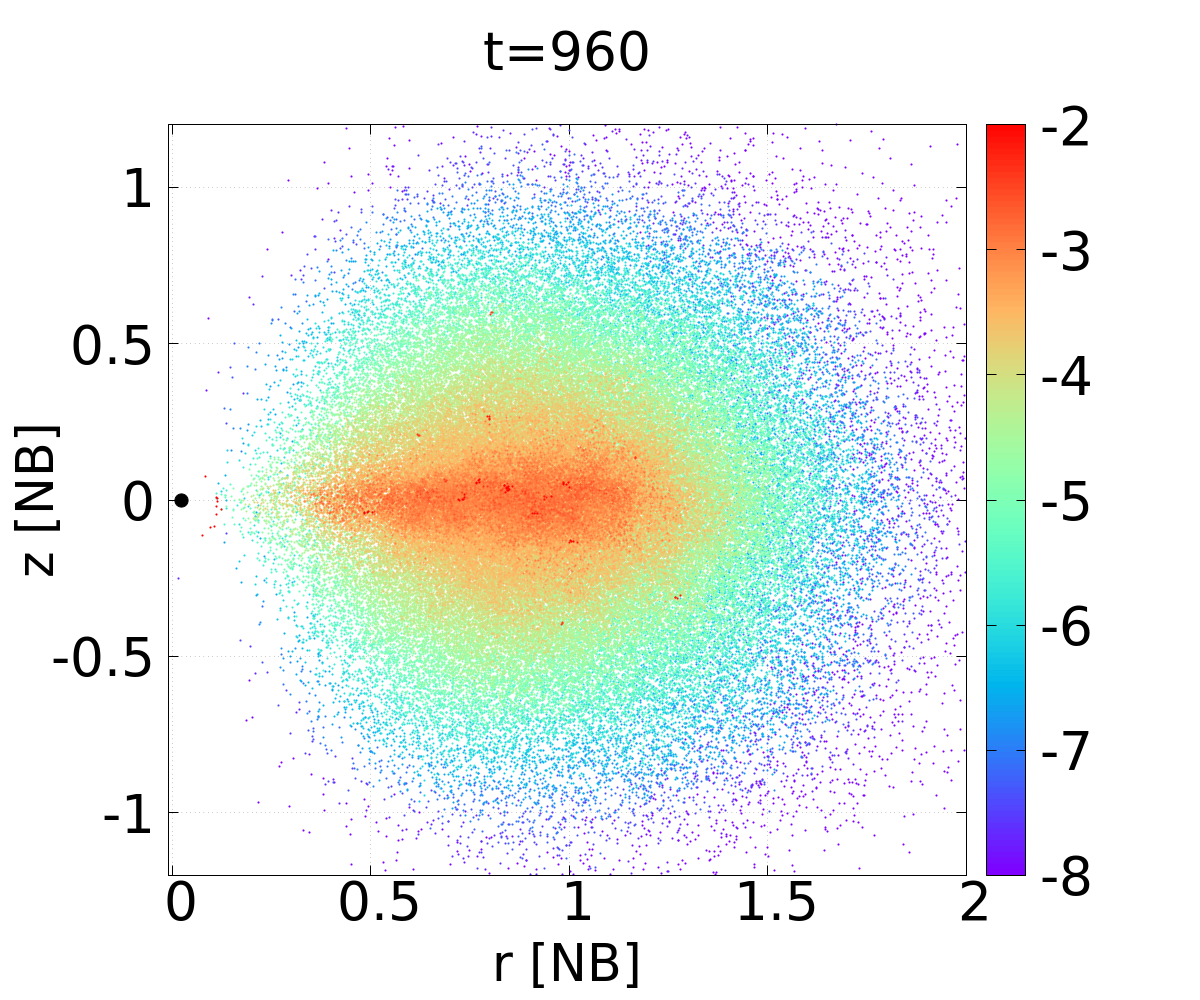}
\caption{Density distribution for the run 0.5-60-0.3. 
Top panels show the face-on orientation and correspond to two consecutive time moments. The bottom panels show the torus in the edge-on orientation (left) and its equilibrium cross-section (right).  The color scale indicates the logarithm of the density. The black point marks the location of the central mass.}
\label{fig:density_i60}
\end{figure}
Fig.~\ref{fig:density_i60} shows the resulting density distribution of the particles, where the asymmetry of the torus shape is clearly visible in both face-on and edge-on projections. The asymmetry forms naturally without any external perturbation and persists throughout the entire simulation. As a result, the particle distribution along the azimuthal angle is inhomogeneous, indicating the formation of a massive overdensity. This overdensity ($m=1$ mode) forms a coherent phase pattern that moves in the prograde direction relative to the orbital motion. Its period, $T_1$, is much longer than the torus orbital period, $T_{\rm orb}$, which is characteristic of a slow mode (see Table~\ref{Table:modes}). The overdensity remains stable throughout the entire integration time up to $t=1000$, which corresponds to 1000 orbital periods of the torus. The torus remains geometrically thick, as illustrated by the density map in the co-moving reference frame (Fig.~\ref{fig:density_i60}, bottom right)\footnote{Note that the persistence of the torus thickness in the presence of self-gravity was demonstrated in our previous simulations for the case of a low-mass torus \citep{2012MNRAS.424..820B,  Bannikova2021}.}.

\begin{figure}[h!]
\centering
\includegraphics[width = 65mm]{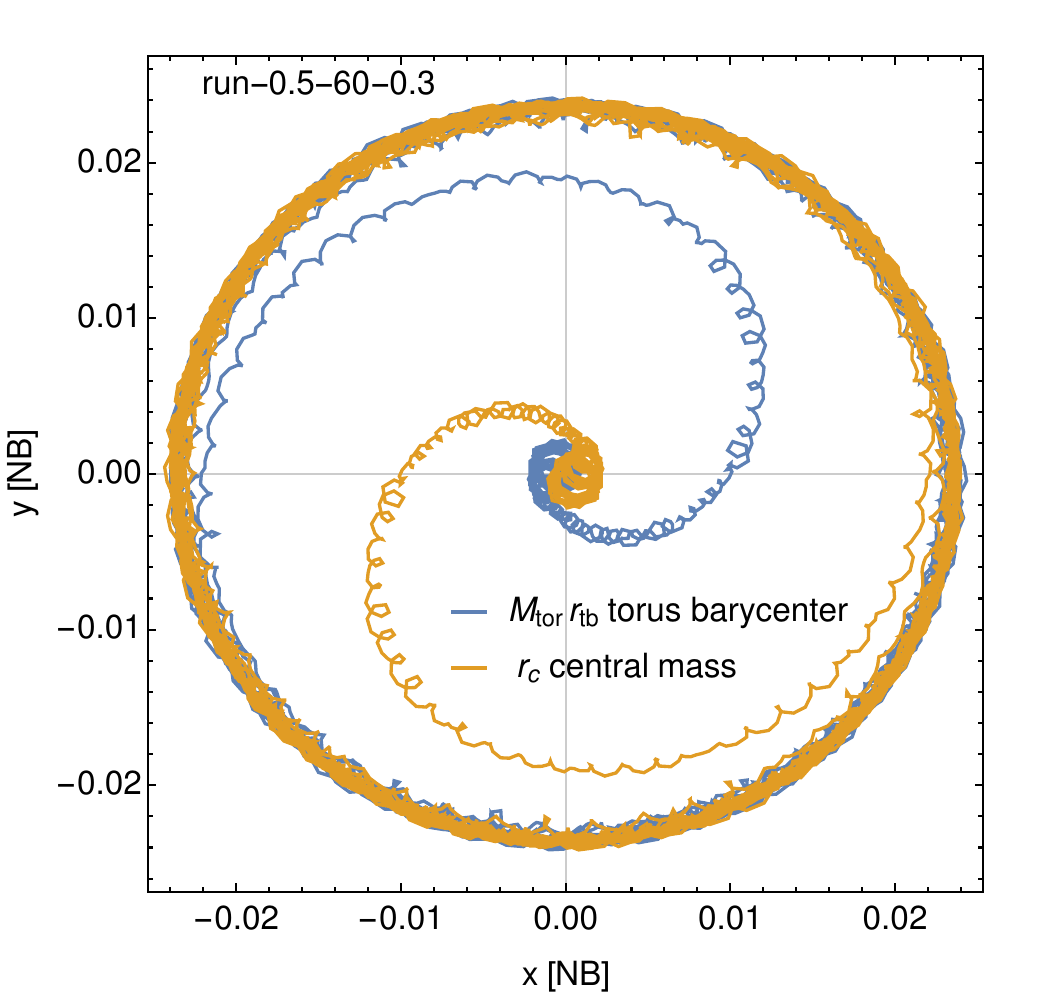}
\caption{Trajectories of $\Mtor \mathbf{r}_\text{tb}$ (blue) and $\Mbh \mathbf{r}_c$ (yellow) over 1000 mean orbital periods of the torus for $\Mtor = 0.1$ (run-0.5-60-0.3).}
\label{fig:torusBar_BH}
\end{figure}

The formation of a large-scale overdensity introduces a global asymmetry in the torus and shifts its barycenter away from the origin, leading to a corresponding motion of the central mass. To quantify this effect, we compute the radius vector of the torus barycenter, ${\bf r}_\text{tb} = (x_\text{tb}, y_\text{tb}, z_\text{tb})$, at each snapshot as the mean position of all particles excluding the central mass:
\begin{equation}
{\bf r}_\text{tb} = \frac{1}{N-1} \sum_{k=2}^{N} {\bf r}_k,
\end{equation}
where ${\bf r}_k$ is the radius vector of the $k$-th particle.
The trajectory of the central mass is in anti-phase with that of the torus barycenter (Fig.~\ref{fig:torusBar_BH}), as required by conservation of the system’s center of mass
${\bf r}_c =- \Mtor {\bf r}_\text{tb}$,
where ${\bf r}_c = (x_c, y_c, z_c)$ is the radius vector of the central mass.
Therefore, the evolution of the torus barycenter can be equivalently traced through the motion of the central mass, which is more straightforward to measure in the simulations.

As seen in Fig.~\ref{fig:torusBar_BH}, the trajectories also exhibit small loop-like features, which likely reflect stochastic particle interactions and the internal evolution of the overdensity. At early times, both the central mass and the torus barycenter undergo oscillations around the origin. As the asymmetric configuration develops, this motion gradually transforms into stable, nearly circular orbits with a constant radius that persists over the entire integration time. This transition marks the establishment of a coherent global $m=1$ structure (see Sect.~\ref{subsec:globalmodes}), with the central mass displacement driven by the internal dynamics of the torus.

\subsection{Relation to virial equilibrium}
\label{subsec:virial}

\begin{figure}[h!]
\centering
\includegraphics[width = 80mm]{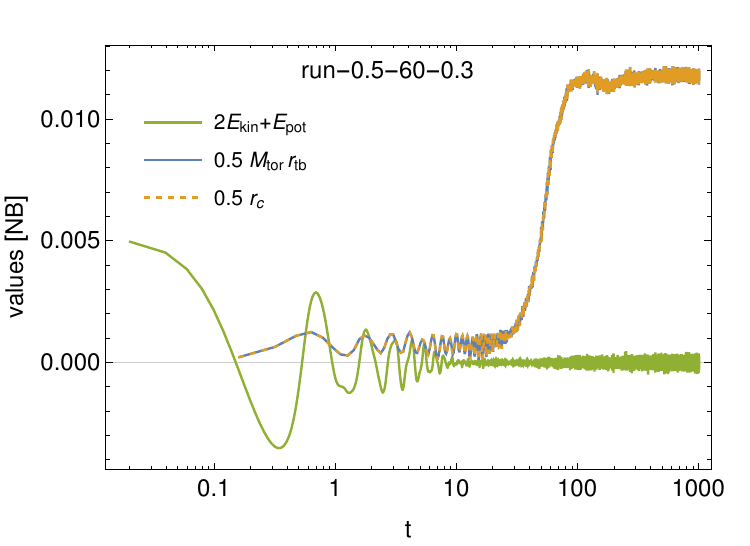}
\caption{The evolution of the virial value (green), radius of the central mass orbit (yellow) which coincides with the curve of the torus barycenter $\Mtor r_\text{tb}$ (blue). The growth of the central mass displacement begins only after virial equilibrium is reached, indicating the spontaneous emergence of the global $m=1$ mode. }
\label{fig:virial}
\end{figure}

The emergence of a stable orbital motion indicates that the system evolves towards a quasi-equilibrium state. To characterize this behavior, we analyze the virial condition of the torus. Fig.~\ref{fig:virial} (green curve) shows the time evolution of the virial quantity \mbox{$2E_\text{kin}+E_\text{pot}$}.

At early times, large-amplitude oscillations are present, reflecting the initial non-virialized state of the system. These oscillations gradually decay, and the system approaches a stationary regime. The orbital radii of the torus barycenter and the central mass begin to increase (Fig.~\ref{fig:virial}, blue and yellow curves) only after the virial quantity has stabilized. This demonstrates that the global 
$m=1$ mode develops only after the system reaches virial equilibrium and is not a transient response to the initial conditions. The same behavior is found in the other runs discussed below, which start from different values, and in some cases different signs, of the virial quantity (see Appendix~\ref{app:D}). The subsequent growth of the asymmetric structure and the associated displacement of the central mass therefore reflect the intrinsic self-gravitating dynamics of the torus. We now examine how this behavior depends on the torus mass.

\subsection{Dependence on the torus mass}
\label{subsec:tormass}

Here we represent the results of the simulations with the same initial particle distribution as for run-0.5-60-0.3 but for the less mass of the torus ($\Mtor = 0.06, 0.01$). 
\begin{figure}[h!]
\centering
\includegraphics[width = 44mm]{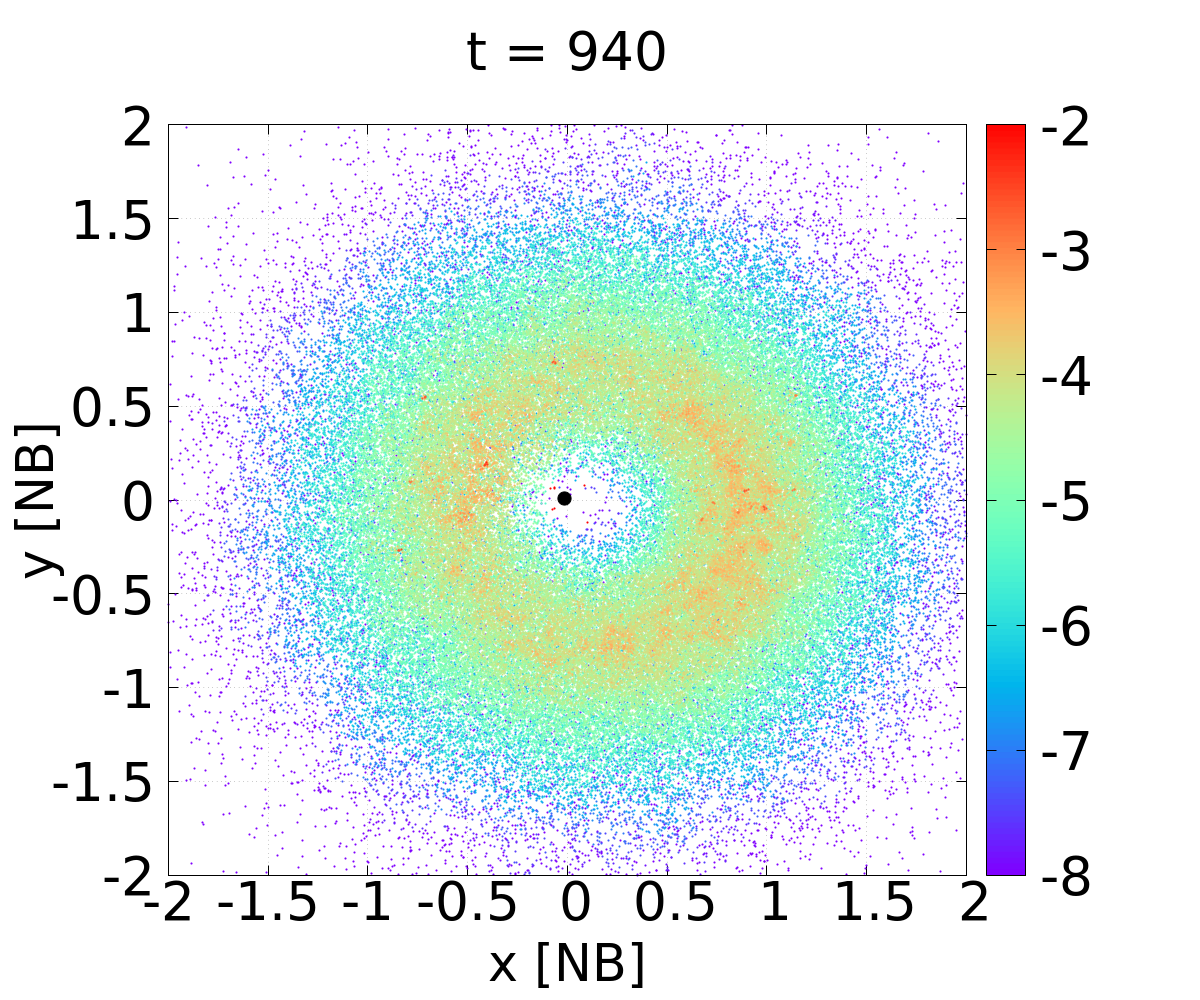}
\includegraphics[width = 44mm]{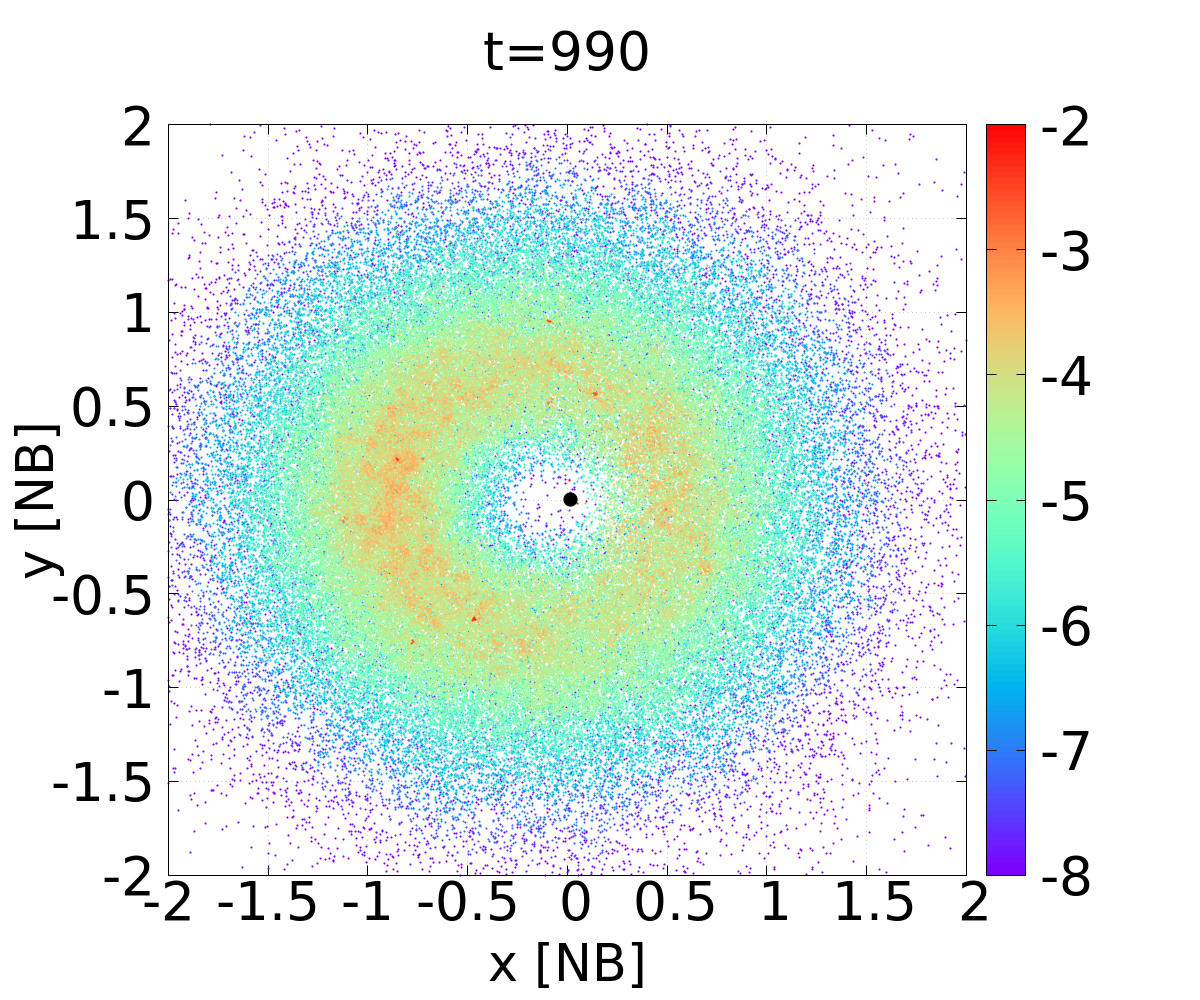}
\includegraphics[width = 44mm]{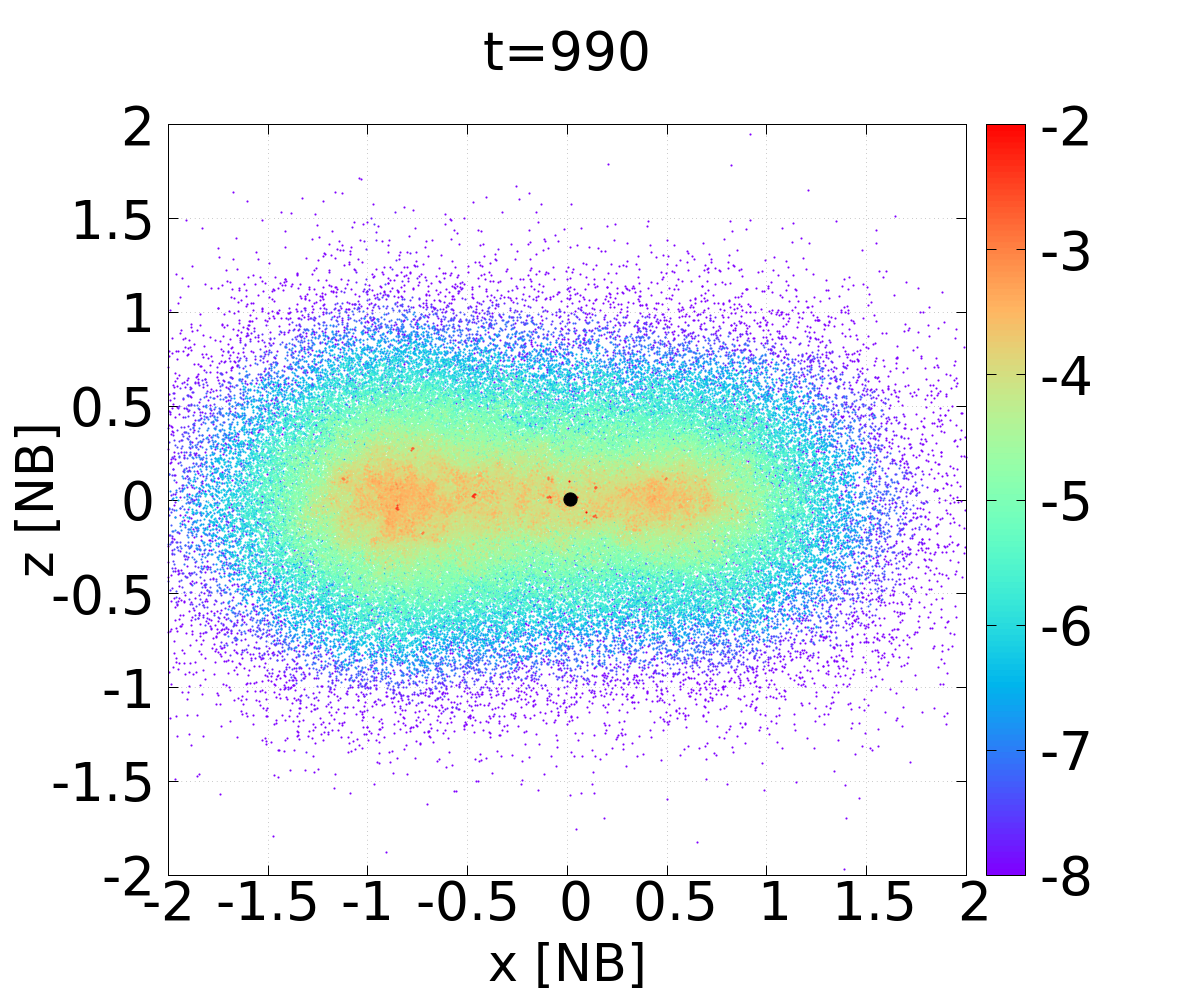}
\includegraphics[width = 44mm]{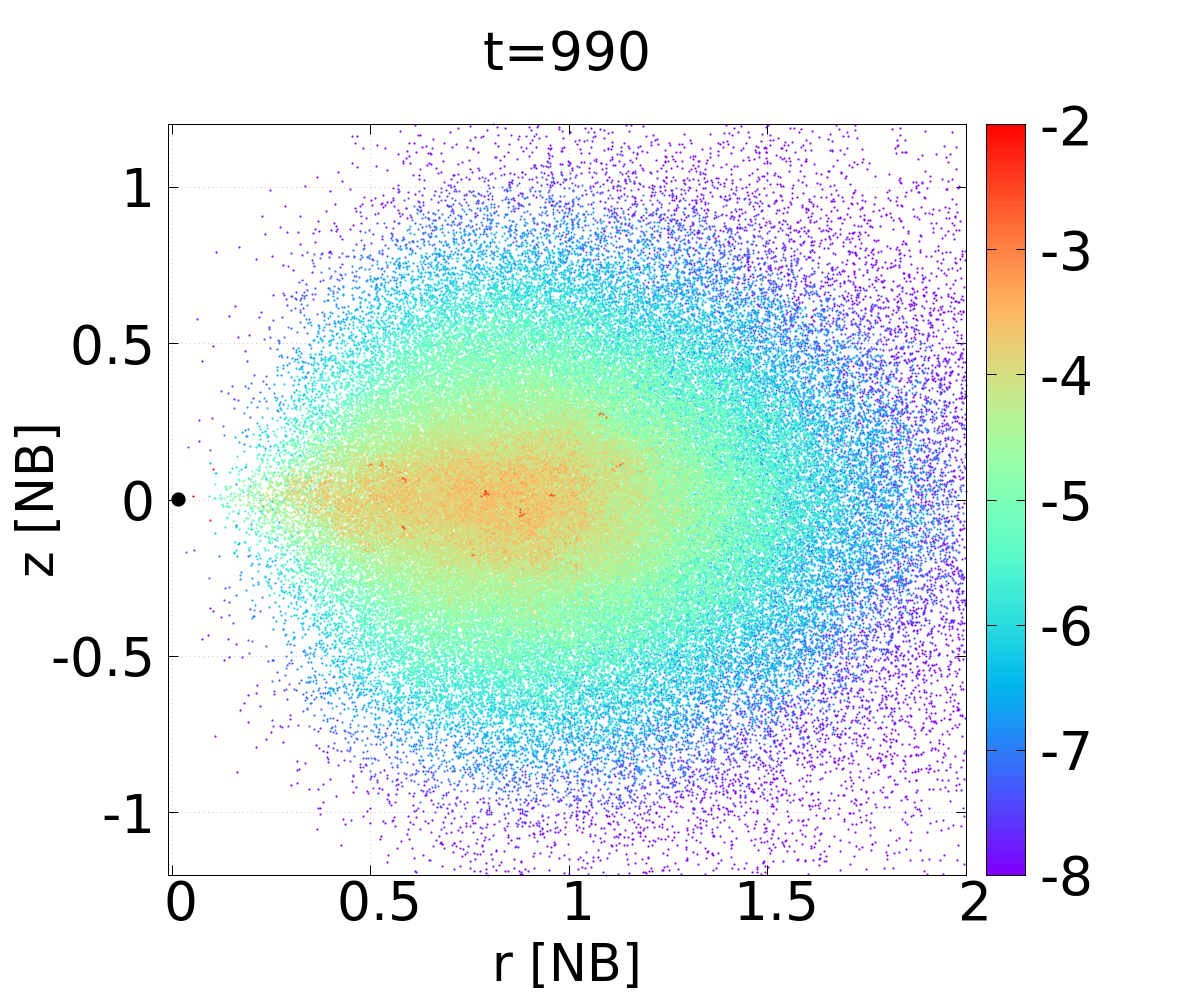}
\caption{ The same that is on Fig.~\ref{fig:density_i60} but for torus mass $\Mtor=0.06$.}
\label{fig:density_i60_m006}
\end{figure}
Fig.~\ref{fig:density_i60_m006} shows clear asymmetry in the torus with $\Mtor = 0.06$ but the density distribution is smoother (compare with Fig.~\ref{fig:density_i60}). The similarity of the torus cross-sections in both cases ($\Mtor = 0.1$ and $0.06$) indicates that the torus remains geometrically thick, with its vertical structure largely independent of the torus mass.
\begin{figure}[h!]
\centering
\includegraphics[width = 80mm]{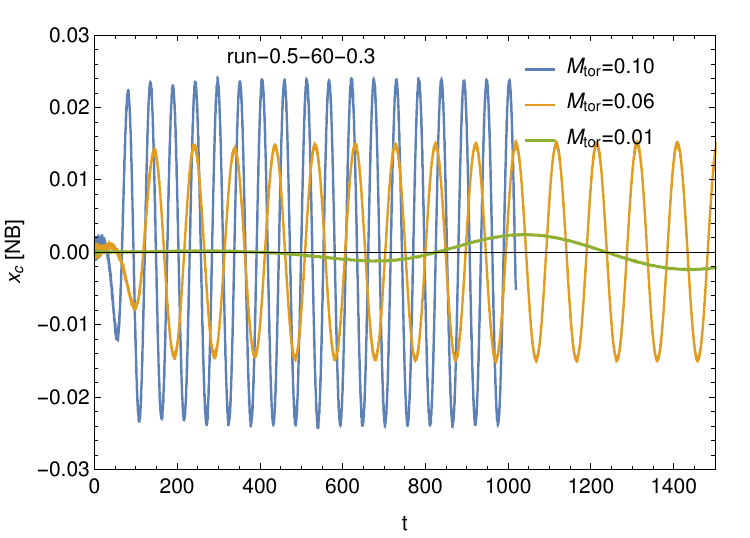}
\caption{Time evolution of the $x$-coordinate of the central mass for tori with masses 0.10 (blue), 0.06 (yellow), and 0.01 (green). All curves correspond to the same initial conditions as for run-0.5-60-0.3. }
\label{fig:diff_mass_05-60-03}
\end{figure}

Fig.~\ref{fig:diff_mass_05-60-03} shows that the coordinate of the central mass decreases approximately linearly with decreasing $\Mtor$, in agreement with the system’s center of mass relation (Sect.~\ref{subsec:form}). Even in the case of the lowest torus mass  ($\Mtor = 0.01$), the  displacement remains clearly detectable, although its amplitude is strongly reduced. This behavior is consistent with an approximately constant number of particles participating in the overdensity for runs with different torus masses but with the same initial distributions. The corresponding differences in the overdensity mass, and hence in the displacement of the central mass, arise from the variation of the particle mass, since $m = \Mtor/N$. 

This self-similar behavior implies that the system’s evolution at different torus masses can be obtained by a simple rescaling of a reference model with identical initial distributions. This significantly reduces the need to simulate a variety of initial distributions for different torus masses; in the following, we therefore perform all runs (Table~\ref{tab:sec1:runs}) for $\Mtor = 0.1$, which provides the strongest and fastest response.

\section{The analysis of perturbation modes}
\label{sec:modes}

To characterize the structure and evolution of the emerging asymmetry, we analyze the perturbation modes of the system.

\subsection{Evolution of the mode amplitudes}
\label{subsec:allmodes}

Consider the particle distribution in the torus in cylindrical coordinates $n = n(r,\phi,z;t)$, where $\phi$ is an azimuthal angle, and $r$ is a radius in the equatorial plane. We first compute the surface density $ \Sigma(r,\phi;t)$ by integrating over the $z$ coordinate of the particles for each snapshot. We then construct a radially averaged azimuthal density profile by averaging the surface density over radius $r$ within the torus body. For this purpose, we divide the torus into $12 \times 100$ sectors ($N_\text{sec}$) in azimuthal angle ($\triangle \psi_\text{sec} = 0.3^\circ$) and compute the number of particles per sector, $ \Sigma(\phi;t)=\langle\Sigma(r,\phi;t)\rangle_\text{r}$. We retain the time dependence by repeating this procedure for each snapshot.

To quantify non-axisymmetric structure in the torus, we perform an azimuthal Fourier decomposition of the radially averaged surface density:
\begin{equation}
    \Sigma(\phi, t) = \Sigma_0(t) + \sum_m \left[ a_m(t) \cos(m \phi) + b_m(t) \sin(m \phi)\right],
\label{eq:density}
\end{equation}
where $a_m$, $b_m$ are the real Fourier coefficients of the $m$-th azimuthal harmonic. The mean number of particles per sector is given by $\Sigma_0=(N-1)/N_\mathrm{sec}$. To estimate the coefficients $a_m$ and $b_m$ for each snapshot, we construct a system of $N_\mathrm{sec}$ linear equations of the form (\ref{eq:density}), including all azimuthal harmonics up to $m=5$. On the left-hand side of each equation, we substitute the measured particle number corresponding to a given azimuthal angle $\phi$, while the right-hand side contains the unknown coefficients $a_m$ and $b_m$  for all modes ($m=1,2,3,4,5$). The resulting overdetermined system is solved using the least-squares method, yielding best-fit estimates of $a_m$ and $b_m$, together with their associated uncertainties, for each mode. The Fourier coefficients can be expressed in terms of the amplitude and the phase: $a_m=A_m \cos\Phi_m$ and $b_m=A_m \sin\Phi_m$. 
Each harmonic can then be written in the equivalent form
\begin{equation}\label{eq:Sigmam}
  \Sigma_m(\phi,t) = A_m(t)\,\cos\big(m\phi - \Phi_m(t)\big), \quad m=1,..
\end{equation}
where $A_m=\sqrt{a_m^2+b_m^2}$ is the amplitude and $\Phi_m=\arctan(b_m/a_m)$ is the phase of the mode. If the azimuthal angle $\phi$ coincides with the phase $\Phi_m$, equation (\ref{eq:Sigmam}) shows that the amplitude $A_m$ gives the particle excess in the direction of the mode maximum.
\begin{figure}[h!]
\centering
\includegraphics[width = 80mm]{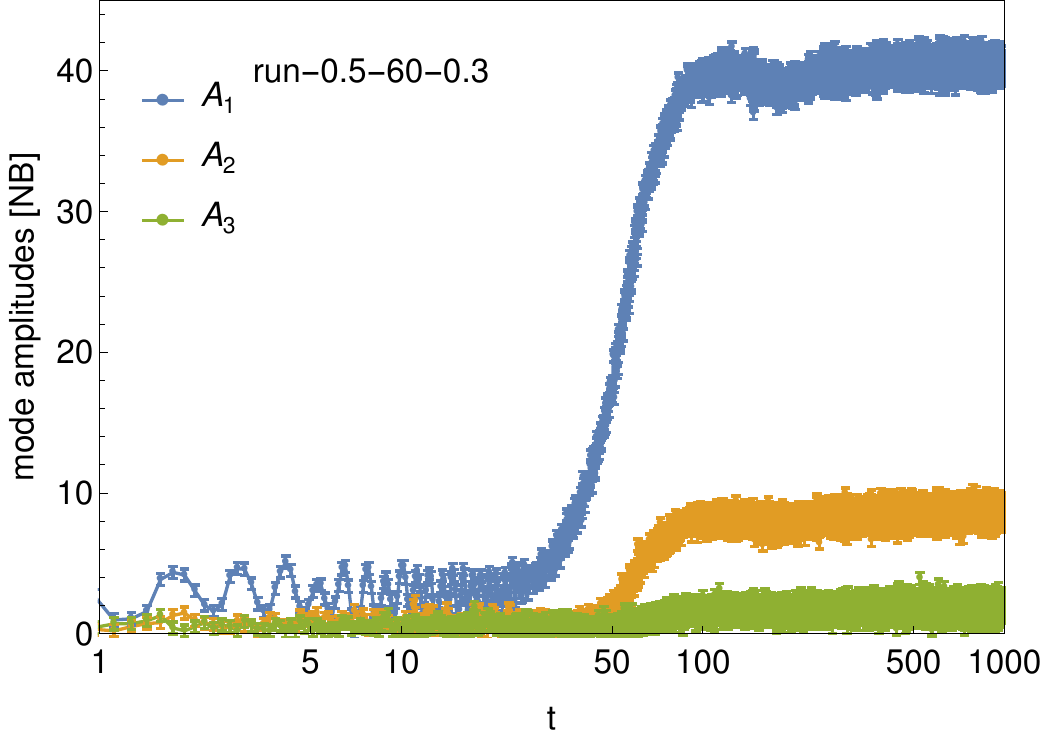}
\caption{Time evolution of the amplitudes of the 
$m=1,2,3$ modes on a logarithmic time scale for the same run shown as a density plot in Fig.~\ref{fig:density_i60}. Error bars indicate the standard deviations.}
\label{fig:modes_05-60-03}
\end{figure}

Fig.~\ref{fig:modes_05-60-03} shows the time evolution of the amplitudes of the first three azimuthal modes, $m=1,2,3$ plotted as a function of logarithmic time for the canonical experiment  run-0.5-60-0.3 (see surface density plot at Fig.~\ref{fig:density_i60}). At early times, all modes exhibit small-amplitude fluctuations associated with the initial relaxation phase. As the system approaches a quasi-stationary state (see Sect.~\ref{subsec:virial}), the amplitude of the  $m=1$ mode ($A_1$) undergoes rapid growth and becomes dominant, reaching a high amplitude which finally remains constant.  
The amplitudes of $m=2$ and $m=3$ modes ($A_2$ and $A_3$) also grow at later times, but remain subdominant throughout the evolution. This behavior indicates that the non-axisymmetric structure observed in the torus is primarily associated with the spontaneous development of a global $m=1$ mode. The $m=4$ mode remains at the noise level and is therefore not dynamically significant in this run.\footnote{The axisymmetric $m=0$ component was also examined for runs with different initial virial states. Its amplitude remains constant within numerical fluctuations and shows no correlation with the growth of the $m=1$ mode.}

\begin{figure}[h!]
\centering
\includegraphics[width = 80mm]{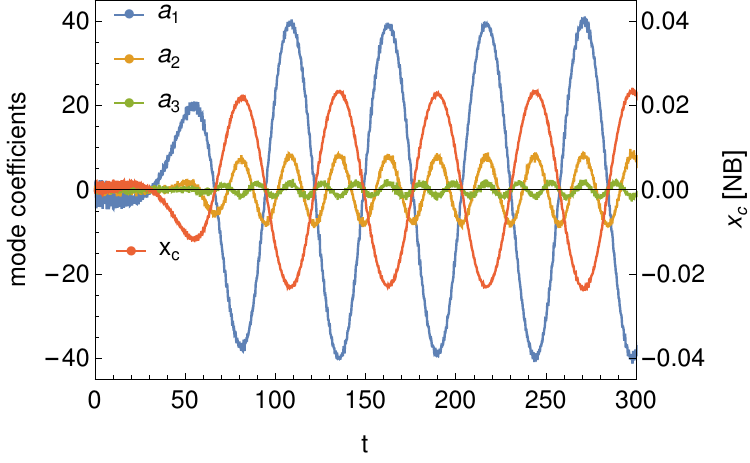}
\caption{Time evolution of the Fourier coefficients of the $m=1,2,3$ modes and the coordinate of the central mass for the same run shown in Fig.~\ref{fig:modes_05-60-03}.}
\label{fig:a123-xc}
\end{figure}

Fig.~\ref{fig:a123-xc} shows the corresponding evolution of the mode coefficients ($a_1,a_2,a_3$) together with the coordinate of the central mass ($x_c$). After an initial relaxation phase, the growth of the $m=1$ mode is clearly correlated with the transition from irregular motion to a coherent orbital motion of the central mass.
Although the $m=2$ and $m=3$ modes remain subdominant, their temporal evolution is correlated both with the $m=1$ mode and with each other. This suggests that higher-order modes are not independent, but are instead dynamically coupled to the dominant $m=1$ structure. 

In the following subsections, we investigate the connection between the non-axisymmetric modes and the orbital elements of the particles. To this end, we convert the Cartesian coordinates and velocities of each particle obtained in our $N$-body simulations into instantaneous (osculating) Keplerian orbital elements using standard two-body relations with respect to the central mass at each snapshot.

\subsection{Global lopsided \textit{m}=1 mode}
\label{subsec:globalmodes}

Here we focus on the connection between the global lopsided 
$m=1$ mode and the distribution of particle eccentricities in the torus. 

\begin{figure}[h!]
\centering
\includegraphics[width = 80mm]{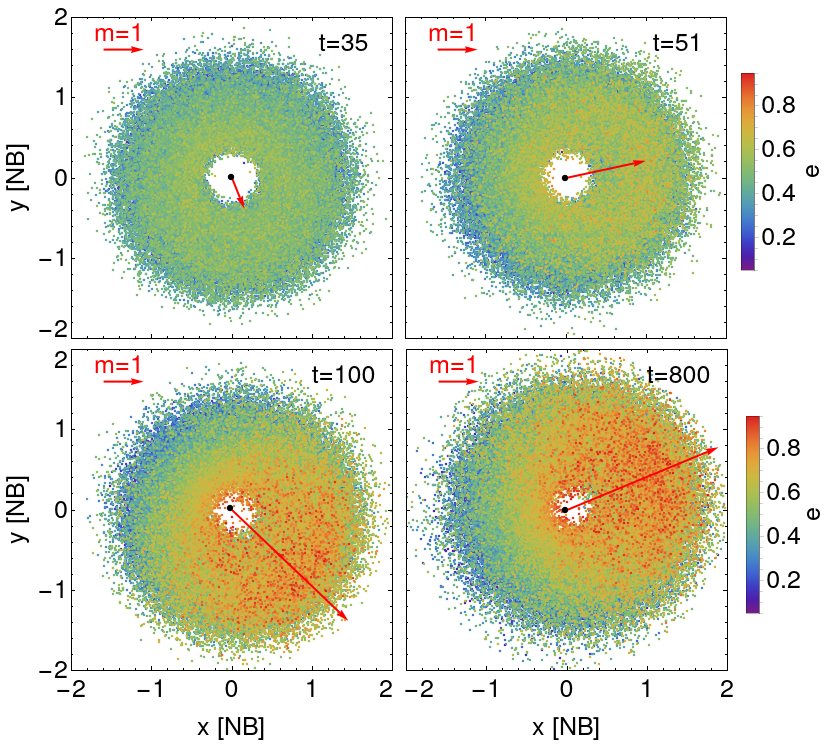}
\caption{Particle distribution in the ($x,y$) plane at four times ($t$=35, 51, 100, 800) for the canonical simulation run-0.5-60-0.3. Colors indicate the eccentricity ($e$) of the particles. The red arrows show the amplitude and phase of the global lopsided $m=1$ mode. }
\label{fig:map_ecc}
\end{figure}
The evolution of the eccentricity distribution in the $(x,y)$ plane at four representative times is shown in Fig.~\ref{fig:map_ecc}; arrows indicate the amplitude and phase of the $m=1$ mode. Initially ($t=35$), the torus is nearly axisymmetric, but as the system evolves ($t=51,100$), a clear lopsided structure develops, with particles reaching higher eccentricities and concentrating on one side, forming a global overdensity aligned with the phase of the $m=1$ mode (Fig.~\ref{fig:map_ecc}).
This asymmetric configuration persists throughout the simulation (e.g. at $t=800$), with the overdensity extending over a wide radial range and dominating the torus morphology. This behavior can be understood in terms of orbital dynamics: eccentric particles spend more time near apocenter, enhancing the density on one side, while the opposite side is depleted as particles pass rapidly through the pericenter. This interpretation is consistent with the mechanism proposed by \citet{1995AJ....110..628T} to explain the double nucleus in M31 (see also Section~\ref{sec:Intr}).
\begin{figure}[h!]
\centering
\includegraphics[width = 85mm]{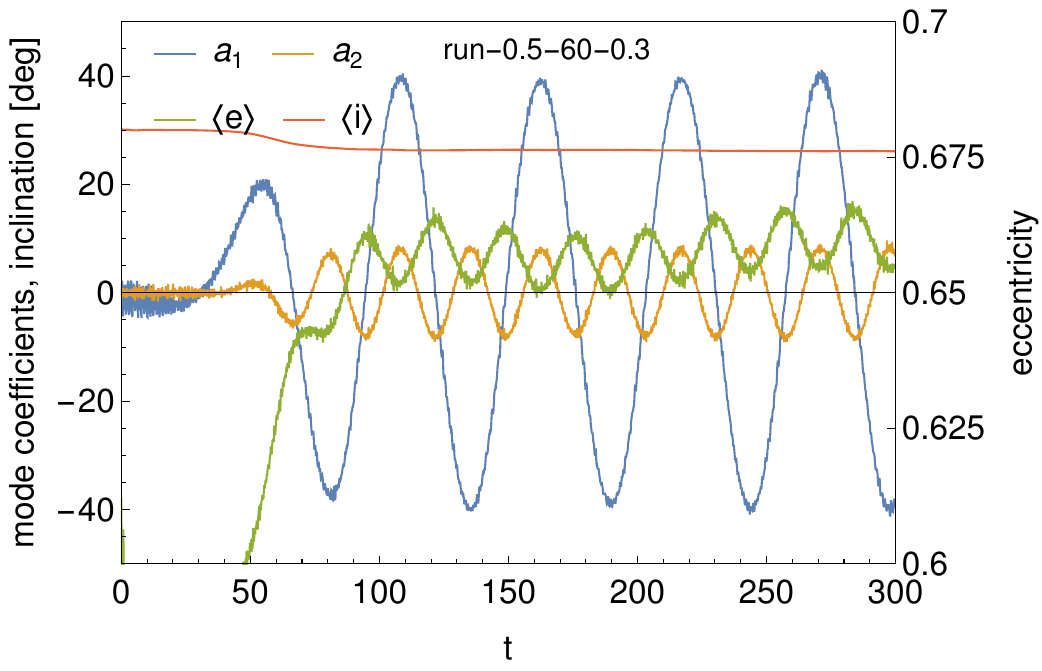}
\caption{ The time evolution of the Fourier coefficients $a_1$, $a_2$, the mean eccentricity ⟨$e$⟩ and the mean inclination ⟨$i$⟩ for the canonical run.}
\label{fig:la1_a2_e}
\end{figure}

The time evolution of the Fourier coefficients together with the mean eccentricity and inclination is shown in Fig.~\ref{fig:la1_a2_e}. The mean eccentricity exhibits small but systematic oscillations in anti-phase with $a_2$. This behavior suggests that the $m=1$ mode is linked to coherent eccentricity variations, while the anti-phase relation indicates a dynamical coupling between the modes. The mean inclination remains nearly constant throughout the simulation, indicating that particles over the full range of inclinations participate in the overdensity (phase pattern).

\subsection{Relation between the orbital elements and the \textit{m}=2,3 modes}
\label{subsec:la1_a2_Omega}

\begin{figure}[h!]
\centering
\includegraphics[width = 80mm]{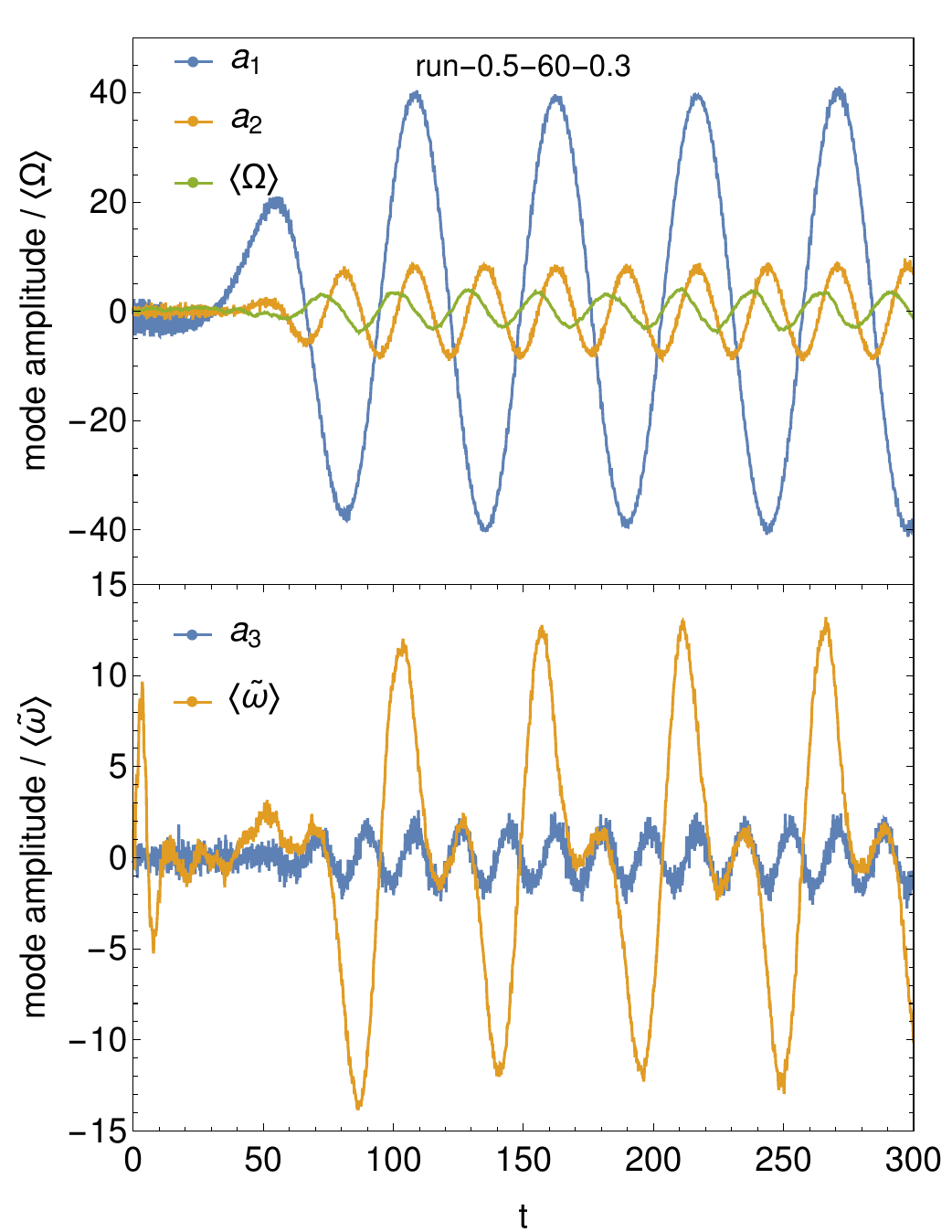}
\caption{Evolution of the Fourier coefficients $a_m$ and orbital elements for the run-0.5-60-0.3. Top panel shows the changing of $a_1$, $a_2$ and the mean longitude of ascending nodes $\langle \Omega \rangle$. Bottom panel shows the same, but for $a_3$ and the mean longitude of periapsis $\langle \tilde{\omega} \rangle$. $\Omega$ and $\tilde{\omega}$ are in degree.}
\label{fig:la1_a2_Omega}
\end{figure}

Fig.~\ref{fig:la1_a2_Omega} (top) shows that the Fourier coefficient $a_2$ oscillates with the same characteristic frequency as the mean longitude of ascending node, $\langle \Omega \rangle$. The bottom panel shows that the evolution of the $m=3$ mode is also related to the mean longitude of periapsis, $\langle \tilde{\omega} \rangle = \langle \Omega_k + \omega_k \rangle$, although the correlation is more complex than in the case of $a_2$ and $\langle \Omega \rangle$.
This behavior suggests that the $m=3$ component is linked to the collective apsidal motion of particles in the torus, rather than arising from random higher-order fluctuations. In this sense, the $m=2$ and $m=3$ modes reflect different aspects of global orbital precession. The $m=3$ component therefore introduces a time-dependent modulation of the azimuthal density structure, which may enhance or suppress particle transport through the overdense region depending on its phase.

To investigate the role of apsidal alignment and its modulation by higher-order harmonics, we construct maps of $\cos(3\Delta \tilde{\omega})$, where
$\Delta \tilde{\omega} = \tilde{\omega}_k - \langle \tilde{\omega} \rangle$, and $\tilde{\omega}_k$ is the longitude of periapsis of the $k$-th particle. These maps illustrate the evolution of the apsidal phase structure and reveal a recurrent transition between differential precession and transient phase alignment associated with the global $m=1$ pattern.

\begin{figure}[h!]
\centering
\includegraphics[width = 80mm]{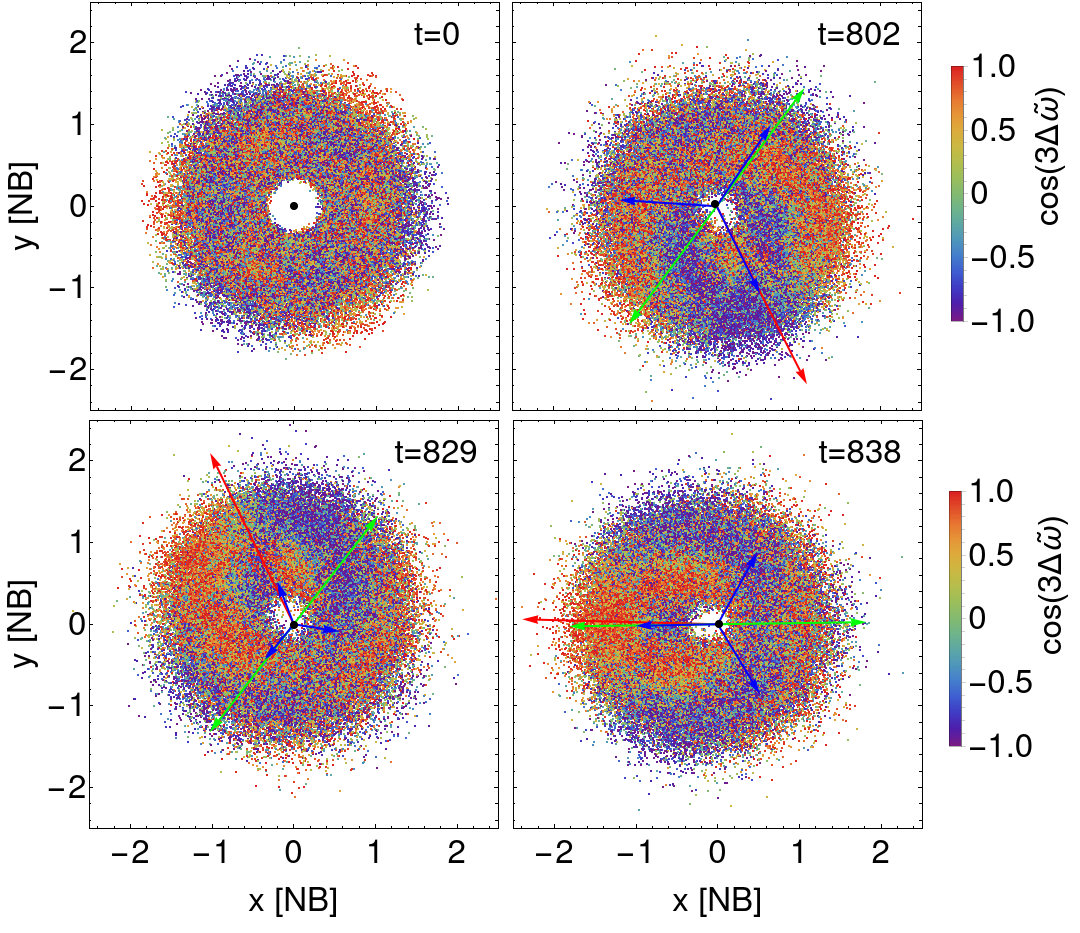}
\caption{Maps of $\cos(3\Delta \tilde{\omega})$ in the $(x,y)$ plane at selected times ($t=$0, 802, 829, 838) for the run-0.5-60-0.3, illustrating recurrent transitions between differential apsidal precession and transient phase alignment with the global $m=1$ pattern. Red, green and blue arrows indicate the instantaneous orientations of the $m=1$, $m=2$ and $m=3$ components, respectively.}
\label{fig:delta}
\end{figure}
At early times ($t=0$), the apsidal angles are distributed symmetrically, indicating the absence of a coherent $m=1$ structure. By $t=802$, a pronounced radial phase gradient develops, reflecting differential apsidal precession: particles at different radii precess at different rates, causing their phases to drift apart and preventing alignment with the overdensity.
By $t=829$, the phase gradient weakens and the system approaches a resonant regime. At $t=838$, a substantial fraction of particles becomes phase-aligned with the $m=1$ mode, forming a coherent structure consistent with transient apsidal synchronization. This alignment is likely related to the self-gravity of the torus, which may promote partial synchronization of apsidal motion. At this stage, the configuration persists for several dynamical times.
The evolution of the $m=3$ component traces the instantaneous phase structure of the system: the threefold pattern directly reflects the distribution of apsidal phases. At later times, coherence weakens again as detuning from resonance restores phase drift, and some particles lose alignment with the pattern. However, the structure is not completely destroyed, but is maintained through a continuous exchange of particles becoming aligned. This sequence repeats over time.

Thus, the observed behavior reflects a competition between differential apsidal precession, which disperses phases, and the self-gravity of the torus, which promotes partial phase synchronization. The long-lived $m=1$ mode is sustained by repeated episodes of this balance. 
This interpretation is supported by simulations without torus self-gravity (see Appendix~\ref{app:C}), where the $m=1$ mode initially develops but subsequently decays, indicating that differential precession alone cannot sustain long-lived phase coherence.

\section{Role of the torus thickness}
\label{sec:thickness}

\begin{figure}[h!]
\centering
\includegraphics[width = 80mm]{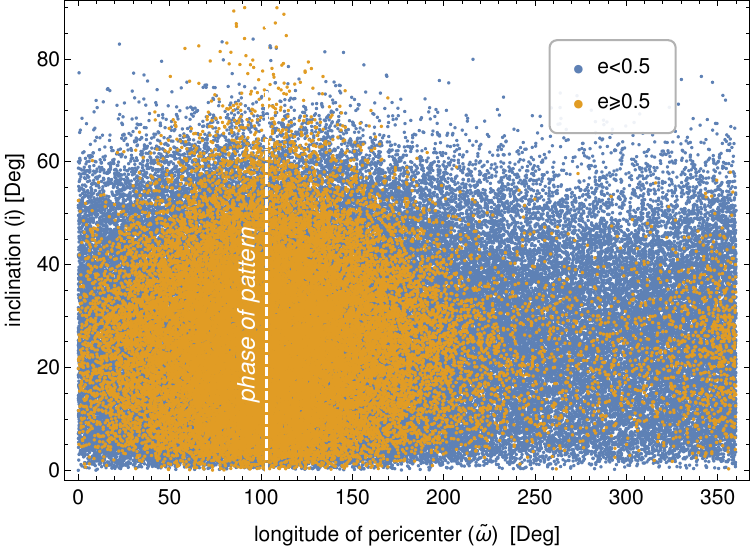}
\caption{The particle distribution in the  $(i,\tilde{\omega})$ plane for the canonical run-0.5-60-0.3 (see Fig.~\ref{fig:density_i60}) at $t=1000$. We choose as an example the particles with $e\ge 0.5$ (orange points) that are participate in  overdensity.}
\label{fig:i-omw}
\end{figure}

To clarify the role of vertical thickness, Fig.~\ref{fig:i-omw} shows the distribution of particles in the $(i,\tilde{\omega})$ phase space for the canonical run 0.5-60-0.3 discussed above (Sect.~\ref{sec:modes}). Particles with $e \ge e_{\max}$ are highlighted in orange, as they correspond to dynamically excited orbits that predominantly populate the overdensity associated with the $m=1$ mode (Sect.~\ref{subsec:globalmodes}). These particles occupy a broad range of inclinations while remaining concentrated in $\tilde{\omega}$ near the phase of the pattern. This shows that the $m=1$ structure is supported by a geometrically thick, three-dimensional population of orbits with coherent apsidal alignment. Consistently, the evolution of individual particle orbits shows coupled variations of eccentricity and inclination, indicating an exchange of angular momentum between radial and vertical degrees of freedom (Appendix~\ref{subsec:ei}).

\begin{figure}[h!]
\centering
\includegraphics[width = 80mm]{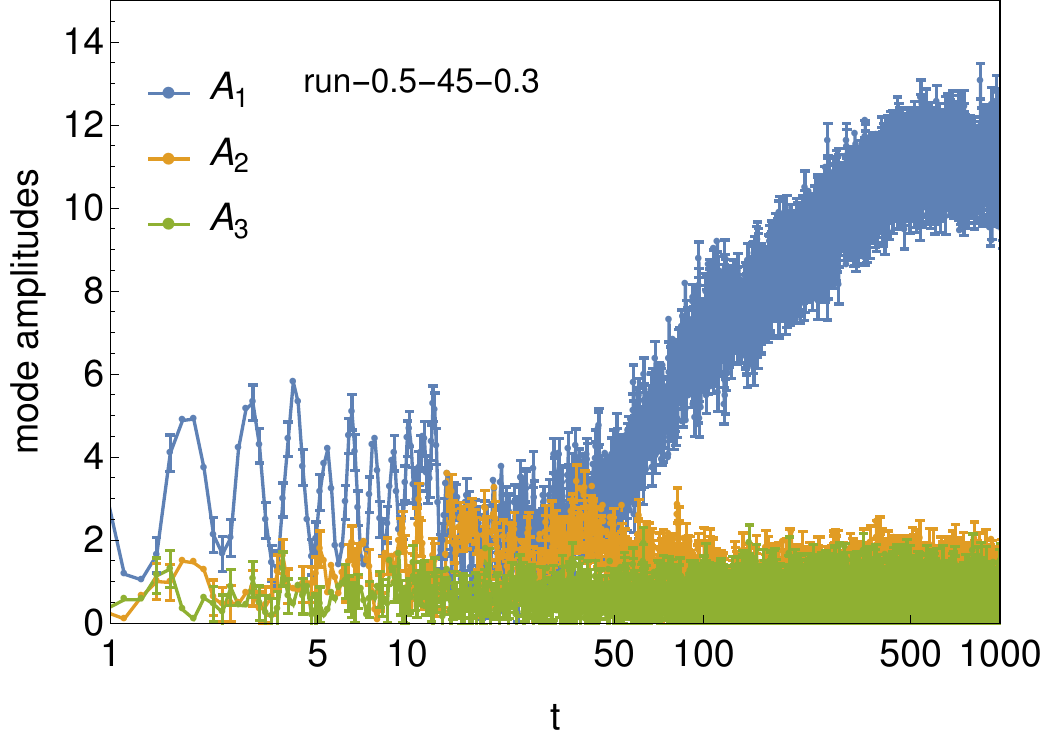}
\caption{The same that is on Fig.~\ref{fig:modes_05-60-03}  but for more thin torus with initial $i_{\max}=45^\circ$  (\mbox{run-0.5-45-0.3}).}
\label{fig:ampl_45}
\end{figure}

We now examine how the initial torus thickness influences the formation and evolution of the non-axisymmetric pattern. To this end, we performed a series of simulations with progressively smaller initial particle inclinations. As a representative example, we consider the case with $i_{\max}=45^\circ$ (run-0.5-45-0.3; Table~\ref{tab:sec1:runs}). Fig.~\ref{fig:ampl_45} shows the evolution of the mode amplitudes for this run. Although all modes exhibit initial fluctuations, only the $m=1$ component shows sustained growth, while the $m=2$ and $m=3$ modes remain at the noise level. In contrast to the thicker torus ($i_{\max}=60^\circ$; Fig.~\ref{fig:modes_05-60-03}), the $m=1$ mode saturates at a significantly lower amplitude, resulting in a much weaker overdensity. For even smaller inclinations ($i_{\max}=30^\circ$ and $10^\circ$), the $m=1$ mode does not develop and remains indistinguishable from numerical noise. This demonstrates that a sufficient vertical thickness is required for both the growth and maintenance of the $m=1$ mode. A thicker torus provides a broader range of orbital configurations and additional degrees of freedom for orbital reorientation, facilitating partial phase synchronization with the overdensity. In thinner configurations, this synchronization is suppressed, preventing the formation of a long-lived lopsided structure.

\section{Dependence on the initial eccentricities and semi-major axes}
\label{Sec:ea}

Other parameters that can influence the final pattern formation are the parameters defining the initial orbital distribution, in particular the maximal eccentricity ($e_\text{max}$) and the characteristic radius ($R_0$),  which determines the range of the semi-major axis spread. 

\begin{figure}[h!]
\centering
\includegraphics[width = 44mm]{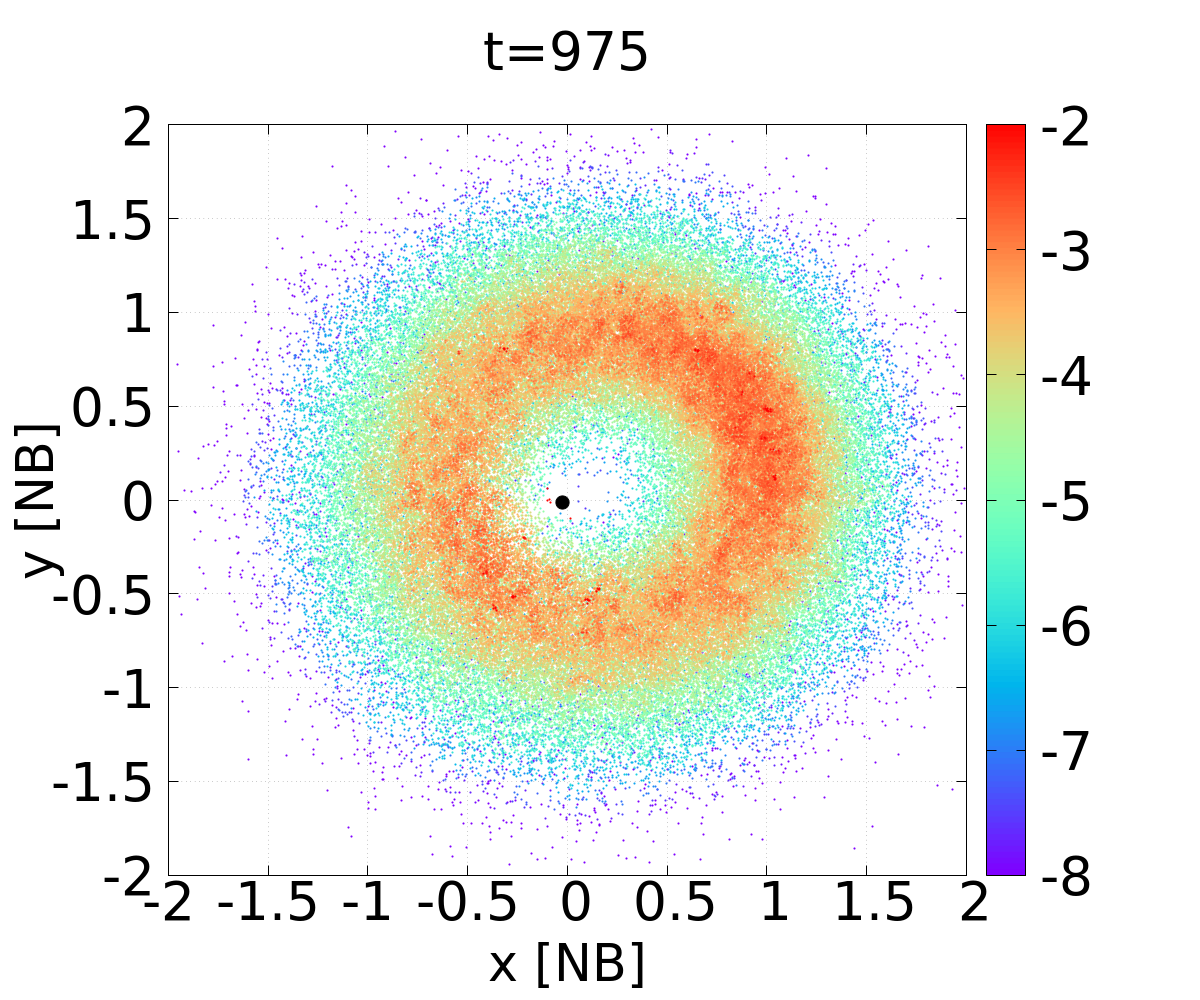}
\includegraphics[width = 44mm]{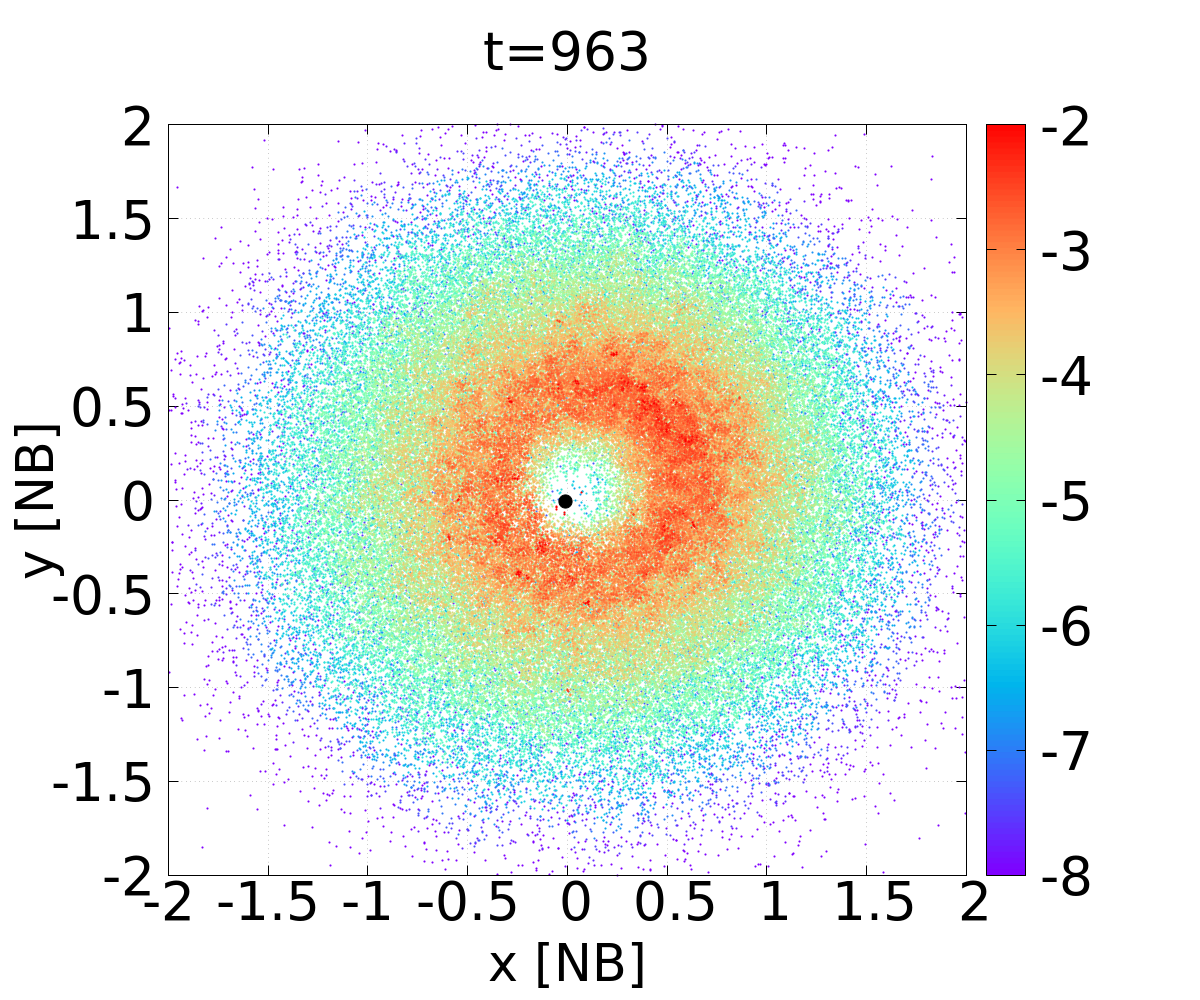}
\includegraphics[width = 44mm]{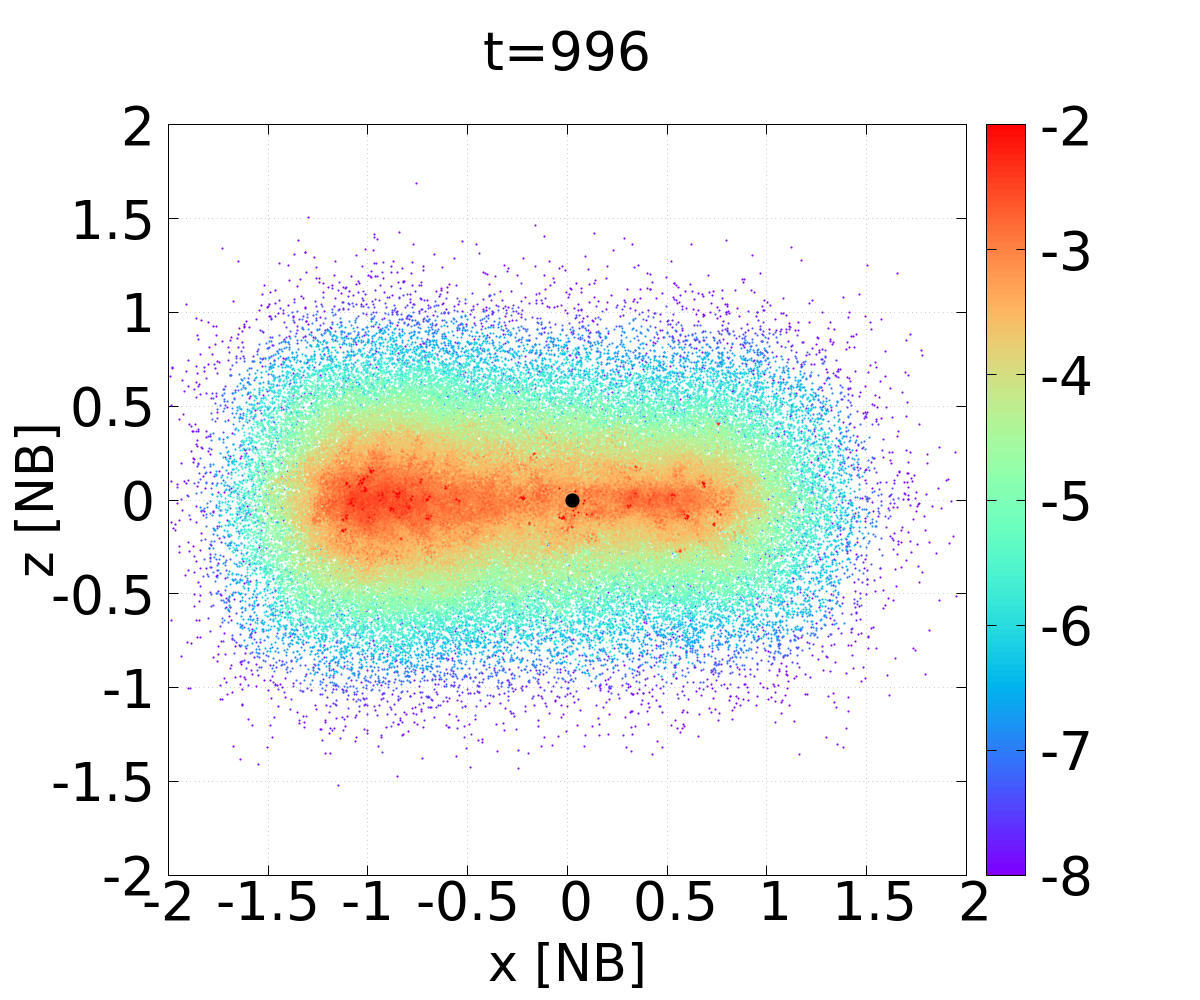}
\includegraphics[width = 44mm]{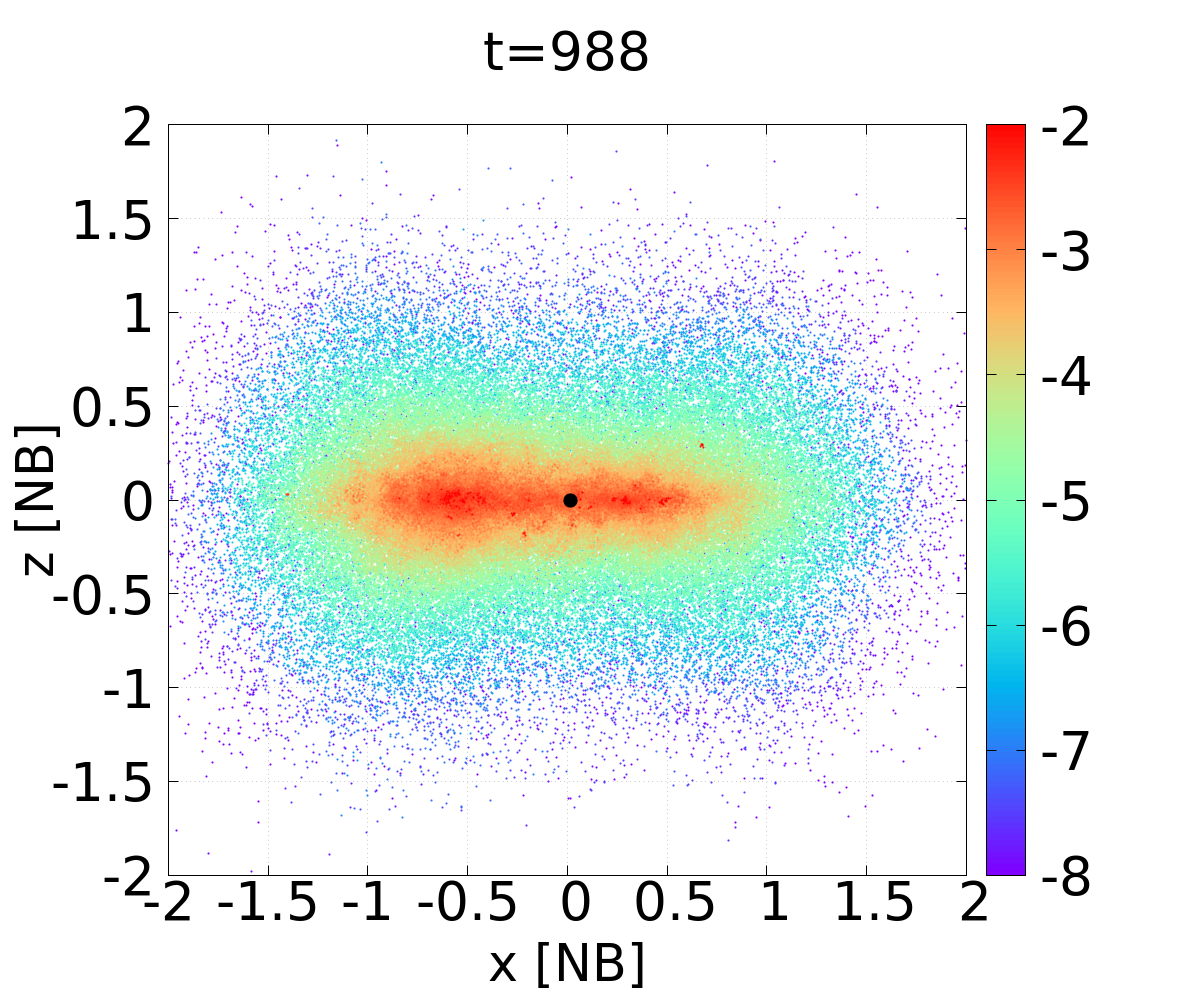}
\caption{The density distribution for two runs with the different initial conditions. Left column corresponds to  run-0.2-60-0.3, right column -- run-0.5-60-0.6.}
\label{fig:density_i60_m01_e02}
\end{figure}

To examine the role of the eccentricity dispersion, we performed a simulation with a smaller initial maximal eccentricity, $e_\text{max} = 0.2$, corresponding to  run-0.2-60-0.3 in Table~\ref{tab:sec1:runs}. In this case, the initial eccentricity distribution is narrower, and the particle orbits are initially closer to circular. The corresponding density plot is shown in Fig.~\ref{fig:density_i60_m01_e02}  (left column). The equilibrium density distribution exhibits a clear large-scale asymmetry related to $m=1$ mode, which is even more pronounced than in the case  run-0.5-60-0.3 with a broader ($e_\text{max}=0.5$) initial eccentricity distribution (see Fig.~\ref{fig:density_i60}). 

Additional simulations were performed for a larger initial spread in the semi-major axis, corresponding to $R_0=0.6$ (run-0.5-60-0.6 in Table~\ref{tab:sec1:runs}). The corresponding density plot is shown in Fig.~\ref{fig:density_i60_m01_e02} (right column), where a clear asymmetry is also present, although the radius of the inner dense region is smaller than in the canonical run. In this case, the particle density increases toward the center due to the wider radial distribution of orbits. 

\begin{figure}[h!]
\centering
\includegraphics[width = 65mm]{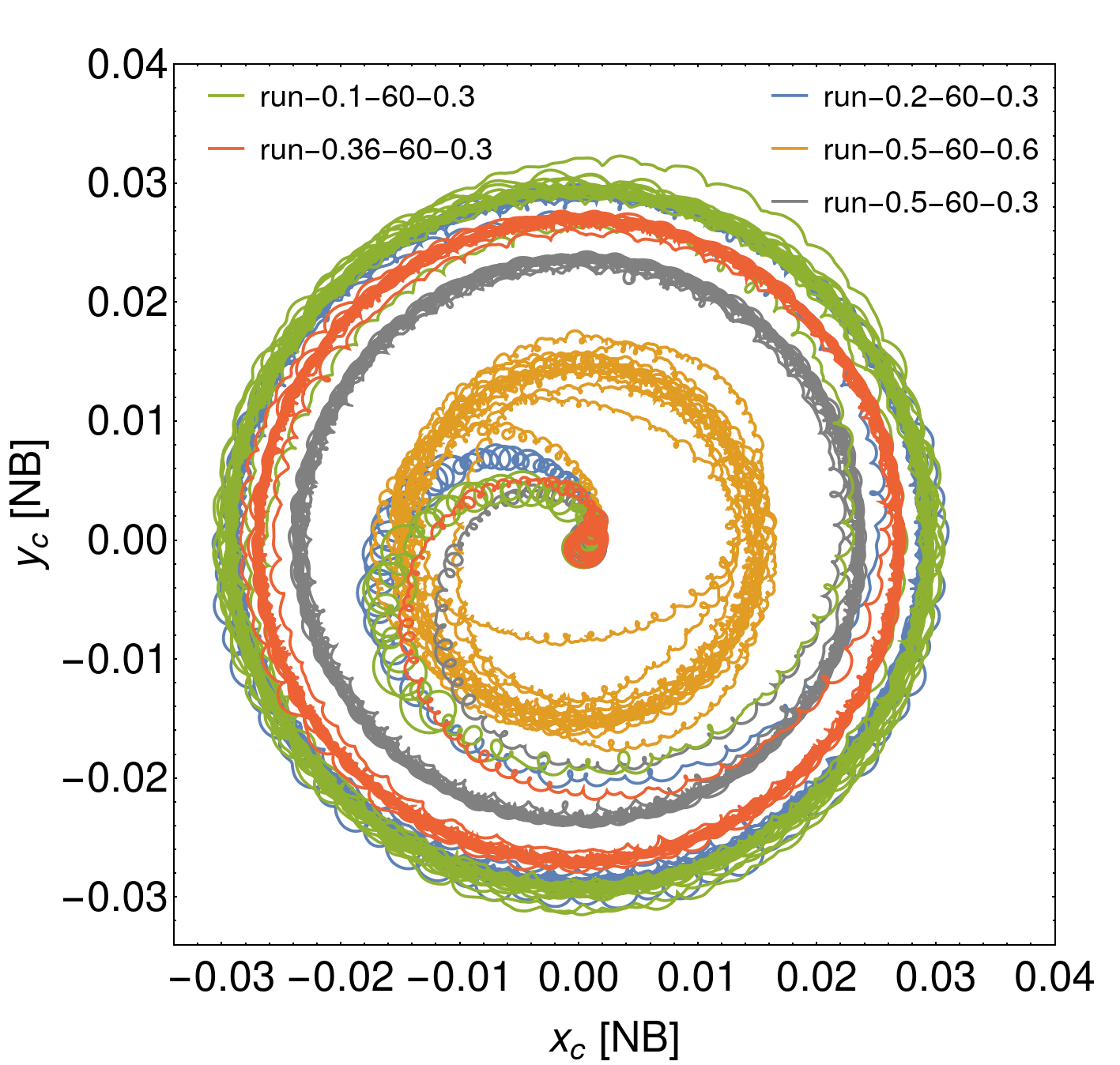}
\caption{The trajectory of the central mass for the run with more circular initial orbits ($e_\text{max}=0.1,0.2,0.3643$ -- green, blue, and orange curves respectively) and for the run with a larger initial semi-major axis range ($R_0=0.6$ -- yellow). The canonical case (run-0.5-60-0.3) is shown in gray for comparison, see Fig.~\ref{fig:modes_05-60-03}.}
\label{fig:mc_orb_ea}
\end{figure}

\begin{figure}[h!]
\centering
\includegraphics[width = 70mm]{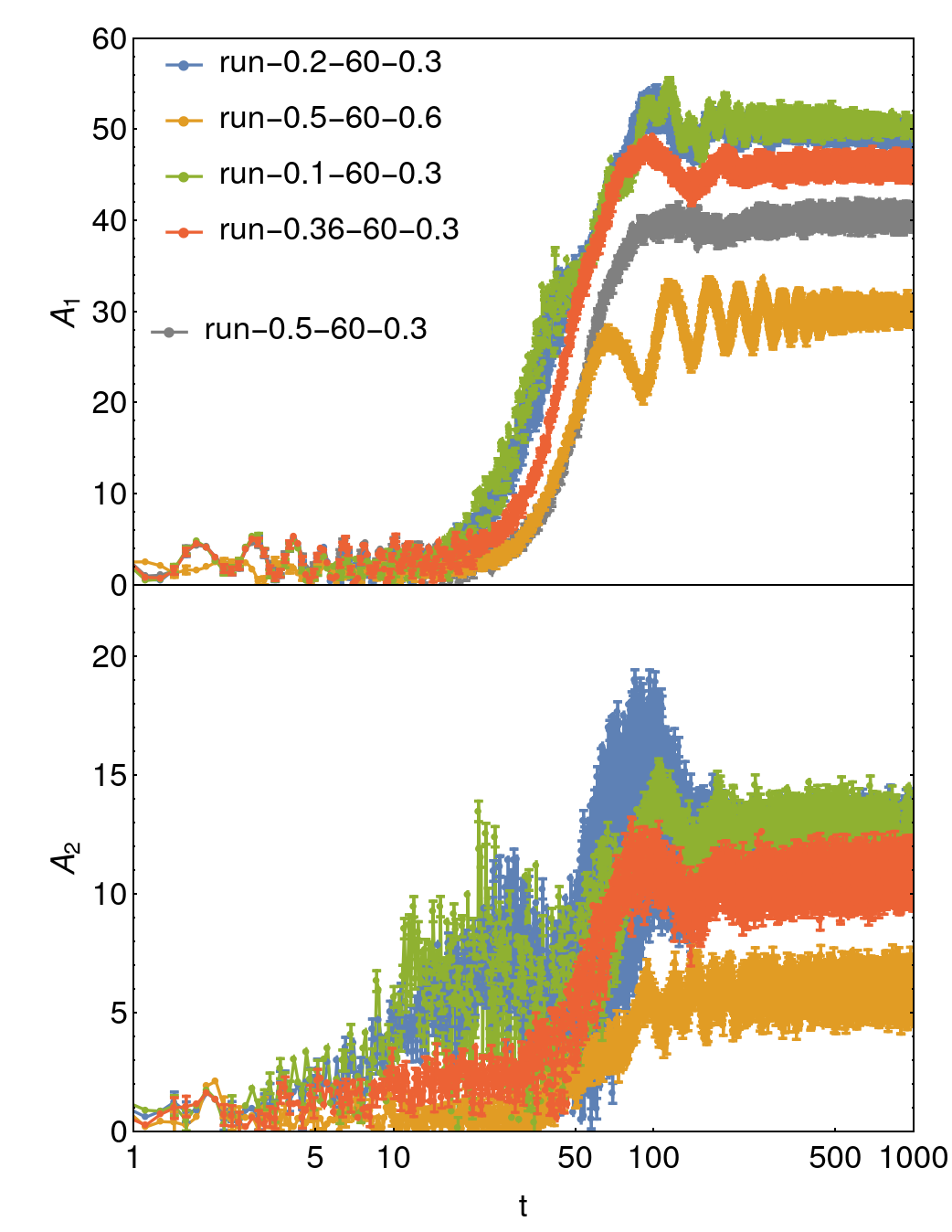}
\caption{Evolution of mode amplitudes for the runs corresponding to Fig.~\ref{fig:mc_orb_ea}. Top panel shows $A_1$ and bottom -- $A_2$ amplitudes.}
\label{fig:A12-ecc_a}
\end{figure}

Fig.~\ref{fig:mc_orb_ea} and Fig.~\ref{fig:A12-ecc_a} illustrate how the initial eccentricity distribution and radial extent of the torus affect the strength and evolution of the non-axisymmetric pattern. A smaller initial eccentricity dispersion ($e_{\max}=0.2$) leads to a faster growth and higher saturation amplitude of the $m=1$ mode, accompanied by a larger displacement of the central mass. This indicates that a more coherent initial orbital configuration facilitates the development of a strong global asymmetry. In this case, the $m=3$ component remains significant, at a level comparable to that in the canonical run, indicating that the presence of higher-order structure is a robust feature of the nonlinear state. In addition, this run shows signatures of $m=4$ component, which is not present in the canonical configuration, pointing to a more strongly nonlinear regime. In contrast, increasing the radial spread of orbits ($R_0=0.6$) results in a weaker $m=1$ mode and a smaller central displacement, indicating that a broader distribution of semi-major axes reduces the coherence of the overdensity.  
 This trend is qualitatively consistent with the linear secular analysis of \citet{Tremaine_2001}, in which self-gravity-dominated slow modes are more readily supported in narrow rings, while a broad radial distribution enhances differential apsidal precession and weakens global coherence. Our system, however, is a geometrically thick, fully self-gravitating torus, and a weaker $m=1$ asymmetry remains long-lived even at larger $R_0$ (see Table~\ref{Table:modes}).
 
Interestingly, despite the different initial conditions, all saturated runs converge toward nearly identical eccentricity distributions in the quasi-equilibrium state (Appendix~\ref{app:ecc_distr}), well described by Rayleigh laws with very similar dispersions. This suggests that the stronger overdensity in the $e_{\max}=0.2$ case is not due to a different final distribution, but rather to the larger fraction of particles that undergo significant eccentricity growth during the evolution. In other words, a narrower initial eccentricity range allows a larger fraction of particles to be dynamically driven into the high-eccentricity tail of the distribution and to participate in the coherent $m=1$ structure.
These results indicate that the nonlinear torus loses memory of the initial eccentricity distribution and approaches a universal orbital state. As we show in the next section, this universality is also reflected in the geometric structure of the overdensity, whose harmonic amplitudes follow a simple scaling relation.

\section{Geometric harmonic scaling and the role of the third mode}
\label{geom_scales}

The nonlinear saturated state exhibits a coherent global structure that extends beyond the dominant $m=1$ asymmetry. Rather than representing a single-mode distortion, the overdensity is accompanied by a systematic hierarchy of higher azimuthal harmonics. In this section we show that these harmonics follow a simple geometric scaling law and argue that this behavior reflects a nonlinear phase-locked configuration of the lowest modes.
 
\subsection{Geometric scaling of harmonic amplitudes}

\begin{figure}[h!]
\centering
\includegraphics[width = 80mm]{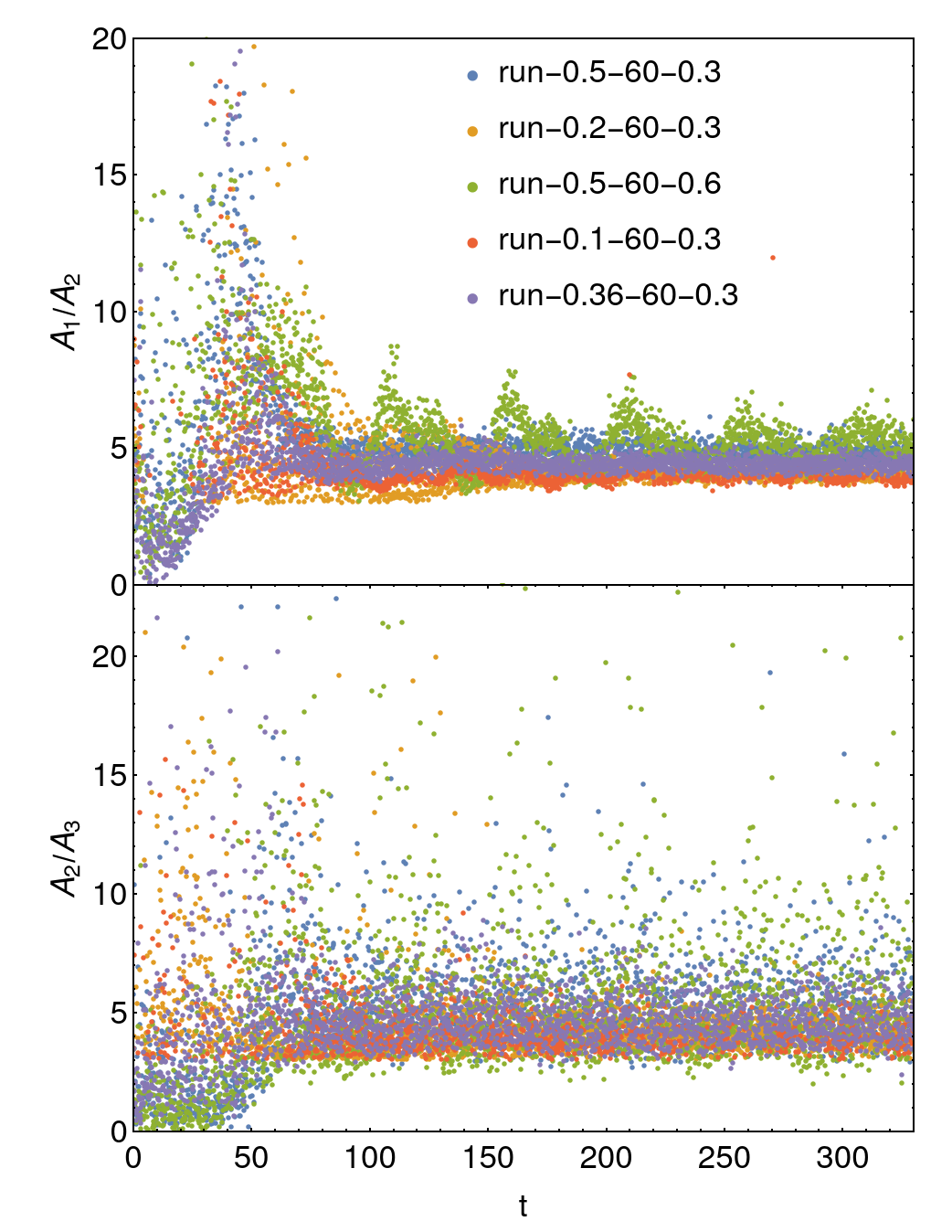}
\caption{Time evolution of the amplitude ratios for different numerical runs corresponding to Fig.~\ref{fig:mc_orb_ea} - \ref{fig:A12-ecc_a}. Top panel shows $A_1/A_2$, bottom -- $A_2/A_3$ ratio.}
\label{fig:RatioA12-ecc_a}
\end{figure}

In the nonlinear saturated state we measure the global azimuthal Fourier amplitudes $A_m$, integrated over the radial extent of the torus (see Sect.~\ref{sec:modes}). Fig.~\ref{fig:RatioA12-ecc_a} shows that, in runs where the $m=3$ mode is present, the amplitude ratios converge to nearly constant values that are only weakly dependent on the initial conditions. Despite different early-time evolution, the ratios $A_1/A_2$ and $A_2/A_3$ become similar in the saturated regime, indicating a nearly universal harmonic structure of the overdensity.

In particular, the ratio of the second to the first harmonic, \mbox{$k \equiv A_2/A_1$}, converges to a well-defined value in each run. Higher-order harmonics follow an approximately geometric scaling,
\begin{equation}
A_m \simeq A_1\, k^{\,m-1},
\qquad
\frac{A_{m+1}}{A_m} \simeq k,
\label{eq:modes_law_1}
\end{equation}
indicating that the harmonic spectrum is effectively controlled by a
single parameter. Table~\ref{Table:modes} summarizes the time-averaged amplitudes in all runs. In the geometric-scaling saturated regime, the parameter $k$ consistently lies in the range $0.2$–$0.25$, despite significant differences in the absolute value of $A_1$. This demonstrates that the nonlinear pattern possesses a nearly universal shape once saturation is reached. In these runs the $m=3$ mode is clearly present, and its amplitude agrees well with the geometric prediction $A_3 \approx A_1 k^2$.
\begin{table}
\caption{Average harmonic amplitudes and geometric scaling diagnostics.}
\centering
\begin{tabular}{lcccccc}
\hline
Run & $A_1$ & $A_2$ & $A_3$ & $k$ & $Q_3$ & $T_1$ \\
\hline
\multicolumn{6}{c}{Geometric-scaling saturated regime} \\
\hline
0.5-60-0.3 & 40.2 & 8.5 & 1.8 & 0.21 & 0.99 & 54.1 \\
0.2-60-0.3 & 49.7 & 12.5  & 3.1 & 0.25 & 1.00 & 44.7 \\
0.5-60-0.6 & 29.9 & 5.9  & 1.2 & 0.20 & 1.05 & 69.6 \\
0.5-60-0.9 & 30.4 & 6.3  & 1.3 & 0.21 & 0.98 & 134.6 \\
 0.1-60-0.3 & 50.0 & 12.9  & 3.3 & 0.26 & 0.99 & 45.5 \\
0.36-60-0.3 & 45.9 & 10.6  & 2.4 & 0.23 & 0.99 & 49.3 \\
\hline
\multicolumn{6}{c}{Weak-saturation / low-amplitude regime} \\
\hline
0.8-60-0.3 & 13.1 & 1.3 & -- & 0.10 & -- & --\\
0.5-45-0.3 & 10.4 & 0.8 & -- & 0.08 & -- & --\\
0.5-30-0.3 & 1.1  & --   & -- & --   & -- & --\\
\hline
\end{tabular}
\label{Table:modes}
\end{table}
To quantify the degree of geometric scaling we introduce the dimensionless combination
\begin{equation}
Q_3 \equiv 
\frac{A_3/A_1}{(A_2/A_1)^2}.
\end{equation}
For a purely geometric spectrum $A_m \propto k^m$, one expects $Q_3 = 1$. In all saturated simulations we find $Q_3$ consistent with unity within a few percent (Table~\ref{Table:modes}). This confirms that the third harmonic is not an independent perturbation but part of a coherent, one-parameter harmonic hierarchy.

The geometric scaling suggests that the nonlinear overdensity is not a superposition of independent modes. Instead, the fixed amplitude ratios indicate that the lowest harmonics are dynamically constrained and form a self-consistent configuration. In the next subsection, we interpret this behavior in terms of an analytic profile and nonlinear mode locking.

\subsection{Analytic representation of the azimuthal profile}

To interpret the measured harmonic hierarchy in the saturated regime (see Table~\ref{Table:modes}), we model the azimuthal overdensity in a co-rotating frame as a single-peaked profile of fixed shape:
\begin{equation}
\Sigma(\phi,t) = \Sigma_0 + S(t)\,F(\psi),
\end{equation}
where $S(t)$ is a global amplitude, $\psi= \phi-\phi_0(t)$ and $\phi_0$ defines the instantaneous pattern phase. In this regime, the overdensity is approximately stationary in the co-rotating frame, so that its shape varies only weakly with time, while the evolution is governed primarily by the phase $\phi_0$.
As a minimal one-parameter representation of such a profile, we adopt the Poisson kernel:
\begin{equation}
F(\psi) =
\frac{1-k^2}{1-2k\cos\psi+k^2} - 1 ,
\label{eq:f}
\end{equation}
where $0<k<1$. The subtraction of unity ensures that the average of $F$ over $\psi$ is zero,  so that $\Sigma_0$ represents the axisymmetric component. This choice is not unique, but provides the simplest analytic form reproducing a geometric harmonic spectrum.
Expanding Eq.~(\ref{eq:f}) into a Fourier series yields
\begin{equation}
F(\psi) =
2\sum_{m=1}^{\infty} k^m \cos(m\psi),
\end{equation}
and therefore 
\begin{equation}\label{eq:SigmaF}
\Sigma(\phi,t) = \Sigma_0 + 2S(t)\,\sum_{m=1}^{\infty} k^m \cos \big(m(\phi - \phi_0 (t)\big).
\end{equation}
Comparing (\ref{eq:SigmaF}) with the harmonic representation (\ref{eq:Sigmam}), we obtain a geometric spectrum $A_m = A_1 k^{m-1}$, in agreement with the empirical relation (\ref{eq:modes_law_1}). We also obtain the phase relation $\Phi_m = m \phi_0$, so that $\phi_0(t) = \Phi_1 (t)$. The higher harmonics thus represent nonlinear distortions of a single coherent overdensity.  
In particular, the phase relation $\Phi_2 \simeq 2\Phi_1$ identifies the $m=2$ component as the second harmonic of the slow $m=1$ mode, rather than as an independent bar or two-armed spiral of the type commonly discussed for galactic disks \citep{2009PhR...471...75J}.

The geometric decay can be equivalently written in exponential form, $A_m=A_1 \exp (-(m-1)w)$, providing a convenient parametrization of the harmonic spectrum. In this sense, $w=-\ln k$ can be interpreted as an effective angular width of the overdensity. 
In this interpretation, the third harmonic arises naturally as part of the coherent nonlinear shape: once the profile width $w$ is fixed, the amplitude of the $m=3$ component is determined by the same parameter that controls $m=2$. The emergence of a one-parameter geometric spectrum therefore implies that the nonlinear state is governed by a constrained, self-consistent coupling between the lowest harmonics.

This geometric hierarchy, however, develops only when the system achieves sufficient phase coherence to form a sharply defined single-peaked overdensity. In simulations with reduced inclination spread (e.g. runs 0.5-45-0.3 and 0.5-30-0.3), as well as in models with a larger initial eccentricity dispersion (e.g. run 0.8-60-0.3), the $m=1$ amplitude remains modest and the higher harmonics do not establish a persistent geometric sequence (see Table~\ref{Table:modes}). In these cases the azimuthal profile stays broad, corresponding to a larger effective width $w$ and a nearly sinusoidal distortion. The appearance of the geometric scaling is therefore not automatic, but instead signals the establishment of a dynamically coherent nonlinear configuration. We now examine the mechanism that enables such phase-locked behavior in the saturated regime.

\section{Possible mechanism for nonlinear mode locking}
\label{sec:mechanism}

The phenomenology established in the previous sections reveals several key properties of the evolution: (i) the spontaneous emergence of a global long-lived $m=1$ mode after the system  reaches virial equilibrium (Sect.~\ref{sec:canonical}), (ii) partial apsidal coherence of a substantial fraction of orbits (Sect.~\ref{sec:modes}), (iii) the emergence of a geometric hierarchy of harmonic amplitudes (Sect.~\ref{geom_scales}), and (iv) the requirement of a finite vertical thickness for the instability to develop and saturate (Sect.~\ref{sec:thickness}). We interpret these properties within a unified dynamical framework.

In a near-Keplerian torus, differential precession tends to disperse apsidal phases and suppress large-scale alignment. The appearance of a coherent $m=1$ structure after the decay of initial virial oscillations therefore indicates that the mode arises from an intrinsic dynamical instability of the self-gravitating torus and saturates in a long-lived nonlinear state. Once the system enters the nonlinear regime, the lowest azimuthal harmonics no longer evolve independently. Instead, their amplitudes and phases become mutually constrained, indicating the establishment of a phase-locked configuration. The geometric scaling of harmonic amplitudes (Sect.~\ref{geom_scales}) provides strong evidence for this constraint. The nearly constant ratio $A_{m+1}/A_m$ implies that the azimuthal structure is effectively governed by a single shape parameter, rather than by a superposition of freely evolving modes. Such behavior is consistent with nonlinear mode locking, in which the harmonics adjust their relative phases and amplitudes to maintain a self-consistent global potential.

Within this picture, the $m=1$ component sets the global eccentric orientation of the torus and defines the large-scale overdensity. The $m=2$ harmonic modifies the curvature of the azimuthal profile and contributes to maintaining a self-consistent gravitational field that limits relative phase drift. The $m=3$ component, whose presence is essential in geometrically thick configurations, appears to provide additional dynamical flexibility. By adjusting within the harmonic hierarchy, it may help absorb temporary phase mismatches associated with differential precession and restore apsidal coherence.

The resulting state is therefore not a rigid-body distortion, but a self-organized nonlinear configuration sustained by the coupled evolution of the lowest harmonics. Differential precession is not eliminated; rather, its disruptive effect is balanced by collective gravitational coupling. When the vertical thickness is reduced (Sect.~\ref{sec:thickness}), this coupling weakens, the geometric hierarchy breaks down, and the global $m=1$ mode fails to reach a stable nonlinear amplitude. 

These results suggest that the long-lived lopsided slow mode arises spontaneously from the intrinsic self-gravitating dynamics of a thick torus and is maintained through nonlinear phase locking between the dominant azimuthal harmonics. The overdensity thus represents a dynamically sustained collective state rather than a transient fluctuation or an imposed eccentric configuration.

\section{Application to astrophysical objects}
\label{sec:astro}

In this section we provide order-of-magnitude estimates illustrating how our results can be applied to stellar double nuclei in galaxies and to asymmetric molecular tori in AGNs.

\subsection{Eccentric stellar nuclei in  Andromeda and  NGC~4486B}

The results obtained in our simulations can be qualitatively applied to eccentric nuclear disks observed in the Andromeda Galaxy (M~31) and in NGC~4486B. In both systems, the observed double nucleus is widely interpreted as a manifestation of a global $m=1$ mode responsible for the asymmetric surface brightness distribution.

Taking into account Eqs. (\ref{eq:density}) and (\ref{eq:Sigmam})  up to the $m=1$ term, we obtain:
\begin{equation}
    \Sigma(\phi) \approx \Sigma_0 +  A_1 \cos(\phi - \Phi_1).
    \label{eq:9a}
\end{equation}
We assume that the projected stellar surface density $\Sigma(\phi)$ is proportional to the observed surface brightness, implying a constant mass-to-light ratio. Let  $\Sigma_{P1} = \Sigma (\phi_{P1})$ and $\Sigma_{P2} = \Sigma (\phi_{P2})$ denote the local surface brightness values at the two observed peaks P1 and P2. The corresponding integrated peak fluxes are
 $F_{P1} \approx \Sigma_{P1} S_1$ and $F_{P2} \approx \Sigma_{P2} S_2$, where $S_1$ and $S_2$ are the effective areas of the two peaks. We then define the ratio:
\begin{equation}
    J \equiv \frac{F_{P1}}{F_{P2}}
    = \frac{\Sigma_{P1}S_1}{\Sigma_{P2}S_2}.
    \label{eq:9b}
\end{equation}
Using the surface-brightness difference $ \Delta\mu = \mu_{P1} - \mu_{P2}$
we have
\begin{equation}
    \frac{\Sigma_{P1}}{\Sigma_{P2}} = 10^{-0.4\Delta\mu} \qquad \text{and}
     \qquad J  = 10^{-0.4\Delta\mu}\frac{S_1}{S_2}.
     \label{eq:9c}
\end{equation}
Hence, the observed projected normalized amplitude can be written as
\begin{equation}
    \tilde{A}_{1,\mathrm{obs}}(\phi) \approx
    \frac{J-1}{J+1}, \quad \text{where}
     \quad  \tilde{A}_{1,\mathrm{obs}}(\phi)
    = \frac{A_1}{\Sigma_0}\cos(\phi-\Phi_1).
    \label{eq:9d}
\end{equation}
This expression explicitly shows that the observed asymmetry depends on the relative orientation between the P1--P2 axis and the phase $\Phi_1$ of the global $m=1$ mode.

For the nucleus of M31, the surface-brightness difference between the two peaks is $\Delta\mu \approx -0.3$ ($\mu_{P1} = 13.4$  mag arcsec$^{-2}$ $\mu_{P2} = 13.7$  mag arcsec$^{-2}$ \citep{2003ApJ...599..237P}), which gives $\Sigma_{P1}/\Sigma_{P2} \approx  1.3$. However, the two peaks differ not only in surface brightness but also in their spatial extent. The P1 component is significantly more extended than P2, suggesting a typical area ratio $S_1/S_2 \sim 7 - 10$ (see Fig.~2 in \citep{2005ApJ...631..280B}). Taking this into account, we obtain $J \sim 4 - 6$,  which yields a normalized amplitude $\tilde A_{1,\mathrm{obs}} \sim 0.8 - 0.86$. Thus, even a relatively modest local surface-brightness contrast corresponds to a much stronger global $m=1$ asymmetry once the geometrical extent of the overdensity is taken into account.

We now apply the same formalism to NGC~4486B. High-resolution observations reveal a double nucleus with a projected separation of $\sim 12$ pc \citep{1996ApJ...471L..79L}, while each photometric peak lies at a distance of $d \sim 6$ pc from the large-scale isophotal center. The eccentric nuclear structure extends out to a radius $R_\mathrm{out} \simeq 20$ pc \citep{Tahmasebzadeh_2025}. Photometric measurements give central surface-brightness values $\mu_{P1} = 12.89$ and $\mu_{P2} = 13.00$ mag arcsec$^{-2}$ \citep{1996ApJ...471L..79L}, implying $\Delta\mu \approx -0.11$ and therefore $\Sigma_{P1}/\Sigma_{P2} \approx  1.1$. Unlike in M~31, the photometric contrast between the two peaks is relatively weak. Adopting $S_1/S_2 \sim 2$ gives $J \sim 2$ and $\tilde A_{1,\mathrm{obs}} \sim 0.38 $. 
Alternatively, an independent geometric estimate based on the displacement of the peaks yields $ \tilde A_{1,\mathrm{obs}} \sim d/R_\mathrm{out} \sim 0.3$, which is of the same order as the value obtained from (\ref{eq:9a}) -(\ref{eq:9d}). 

These results demonstrate that the proposed framework can account for the observed asymmetries in eccentric stellar disks in galactic nuclei, even within the limitations of simplified photometric estimates. The inferred amplitudes depend on projection effects, geometry, and disk thickness, and a complete description requires incorporating kinematic constraints and performing dedicated simulations for each object, which will be addressed in a forthcoming study.

\subsection{Molecular torus in Seyfert galaxy NGC~613}

Recent ALMA observations of the nearby galaxy NGC\,613 hosting an AGN reveal that the molecular torus exhibits a pronounced non-axisymmetric structure. The torus has a characteristic radius of $\Rtor \simeq 8$ pc, while the depleted central region (“hole”) is displaced by about $d \simeq 3$ pc from the position of the AGN. Such an offset indicates a lopsided distribution of the molecular gas and has been interpreted as possible evidence for an $m=1$ asymmetry in the torus \citep{2026A&A...705A.124C}. 

The estimated torus-to-SMBH mass ratio in this object is $M_{\rm tor}/M_{\rm BH} \sim 0.3$, indicating that the self-gravity of the torus may play a dynamically significant role. The observed geometrical offset cannot be directly identified with the Fourier coefficient of the $m=1$ mode, but it provides a useful order-of-magnitude estimate of the non-axisymmetric distortion. We assume that the displaced central cavity approximately traces the barycenter of the torus. For a rough estimate, we use the result of our canonical experiment run-0.5-60-0.3, where $\Mtor=0.1$. The displacement of the torus barycenter is $r_\text{tb} = 0.24\Rtor$ and the amplitude of the $m=1$ mode is $A_1 \approx 40$. Assuming an approximate linear scaling of the displacement with torus mass (see Sect.~\ref{subsec:tormass}), for $\Mtor = 0.3$ we obtain $r_\text{tb} = 0.72\Rtor$. Using the torus radius from observations, we have $r_\text{tb} = 5.76$~pc. This distance is larger than the observed value of 3~pc. This estimate should therefore be regarded as an upper limit. From another side, we can suggest that the actual amplitude of the asymmetry is lower, $A_1 \approx 20$ which corresponds to the 3~pc depleted region. This discrepancy may be due to dissipation effects between the clouds that reduce the effective asymmetry. We plan to investigate this in future work.  

\subsection{Offset of SMBH}

One of the most significant outcomes of our simulations is the displacement of the central mass relative to the system barycenter, which develops as a dynamical response to the formation of a persistent overdensity in the torus. In galactic nuclei, this corresponds to an offset of the supermassive black hole (SMBH). We use for the estimation our canonical run-0.5-60-0.3 for which the offset of the central mass $\Delta r_{\rm c} = \Delta r_{\rm BH} \sim 0.024$. Adopting for M~31 $M_{\rm BH} \approx 10^8 M_\odot$ \citep{2005ApJ...631..280B}, the Schwarzschild radius is $r_s \sim 10^{-5}$~pc, and a nuclear stellar disk mass of $\approx 0.1M_{\rm BH}$ \citep{2001A&A...371..409B}, we scale our dimensionless model to physical units by assuming that one unit of length corresponds to 1~pc. This scaling approximately matches the observed P1–P2 separation in M31 which is $\sim 1.8$~pc \citep{1993AJ....106.1436L}. In our canonical experiment, this implies the SMBH offset $\Delta r_{\rm BH} \sim 2\times 10^3 r_s \sim 0.02$~pc. Taking into account the distance to M31 $D = 785$~kpc \citep{2005MNRAS.356..979M}, it corresponds to the angular offset $\Delta \theta_{\rm BH} \approx 6$~mas.  
A similar order-of-magnitude estimate can be obtained for NGC~4486B. Assuming a characteristic nuclear scale of $R \sim 6$~pc \citep{1996ApJ...471L..79L} yields an expected SMBH offset of $\Delta r_{\rm BH} \sim 0.1$~pc. At a distance of $D = 16$~Mpc \citep{1997ApJ...482L.139K}, this corresponds to an angular displacement $\Delta \theta_{\rm BH} \approx 2$~mas. 

The estimated SMBH offsets for M~31 and NGC~4486B are comparable to current observational resolution limits and may therefore become detectable with future high-sensitivity radio interferometric observations.

\section{Conclusions}

We have investigated the evolution of a three-dimensional, self-gravitating, collisionless torus orbiting a central mass using direct $N$-body simulations. Our primary goal was to determine whether a long-lived global $m=1$ configuration can arise spontaneously from an initially axisymmetric state and to identify the mechanism responsible for its persistence.

Our simulations demonstrate the following main results. 

\begin{itemize}

\item A global lopsided ($m=1$) slow mode forms spontaneously once the torus reaches a quasi-equilibrium state, without any imposed eccentricity or external perturbation. The asymmetry is therefore an intrinsic dynamical outcome of self-gravity in a near-Keplerian torus.

\item The long-lived overdensity is a coherent phase pattern sustained by partial apsidal alignment of a substantial fraction of orbits. The \mbox{$m=1$} pattern speed is set by the collective precession of these orbits and scales nearly linearly with the torus mass. 

\item The persistence of the lopsided structure requires nonlinear coupling of low-order modes. The $m=1$ component defines the global eccentric orientation, while the $m=2$ mode is phase-locked with $m=1$ and limits relative phase drift. The $m=3$ component, excited in sufficiently thick configurations, contributes to maintaining apsidal coherence.

\item A sufficient vertical thickness of the torus is a necessary condition for both the growth and long-term maintenance of the $m=1$ mode, providing the additional degrees of freedom required for phase alignment. Thin configurations fail to develop a coherent asymmetry, indicating that the instability is intrinsically three-dimensional.

\item In the saturated regime, the properties of the nonlinear state show only a weak dependence on the initial conditions. Both the harmonic structure and the eccentricity distribution converge toward nearly universal forms, indicating that the system loses memory of its initial configuration.

\item As a dynamical consequence of the persistent overdensity, the central mass acquires a displacement with respect to the system barycenter. The amplitude of this displacement scales with the torus mass and, when expressed in physical units, may lead to observable offsets of SMBHs in galactic nuclei.

\item The mechanism identified here is consistent with observed asymmetries in both eccentric stellar nuclei and molecular tori in AGNs, suggesting a common dynamical origin. The inferred amplitudes are of the same order as those obtained in our simulations, supporting the interpretation of these systems as manifestations of a global $m=1$ slow mode.

\end{itemize}

Taken together, these results show that eccentric nuclear configurations can arise naturally in geometrically thick, self-gravitating near-Keplerian tori. The mechanism identified here — spontaneous formation of a global $m=1$ slow mode sustained by nonlinear phase locking — provides a dynamical framework for understanding the origin and evolution of double nuclei and other large-scale asymmetries observed in galactic centers.

\begin{acknowledgements}

SS gratefully acknowledges the support under a grant agreement \mbox{$\Gamma$/19-25} between Italian National Institute for Astrophysics (INAF) and V.N.~Karazin Kharkiv National University (Ukraine).
PB and MI gratefully acknowledge the Polish high-performance computing infrastructure PLGrid (HPC Center: ACK Cyfronet AGH) for providing computer facilities and support within computational grant No.~PLG/2026/019243.
The authors thank the anonymous reviewer for helpful corrections, comments, and suggestions.

\end{acknowledgements}

\bibliographystyle{aa}  
\bibliography{references} 

\begin{appendix}
\section{Robustness of the results}

\label{App:A}

For our canonical experiment considered in Sect.~\ref{sec:canonical}-\ref{sec:modes}, we performed additional runs varying numerical parameters (the number of particles $N$ and the softening parameter $\epsilon$), as well as introducing a more homogeneous initial distribution (see Table~\ref{tab:app1:runs}). As a diagnostic of the system evolution, we consider the behavior of the central-mass orbital radius, as it is the most sensitive to initial fluctuations and reflects the global dynamical response of the system.
\begin{table}[htbp!]
\caption{Test parameters of run-0.5-60-0.3.}
\centering
\begin{tabular}{ccccccc}
\hline
Run name &  $N$ & $\epsilon$  \\
\hline
128k;  $\epsilon=10^{-2}$ & 128k & $10^{-2}$    \\
256k & 256k & $10^{-2}$    \\
64k & 64k & $10^{-2}$  \\
$\epsilon=10^{-3}$ & 128k & $10^{-3}$   \\
$\epsilon=10^{-4}$ & 128k & $10^{-4}$     \\
sym & 128k & $10^{-2}$   \\

\hline
\end{tabular}
\tablefoot{The remaining parameters are listed in Table~\ref{tab:sec1:runs} and $\Mtor=0.1$.}
\label{tab:app1:runs}
\end{table} 
\begin{figure}[h!]
\centering
\includegraphics[width = 85mm]{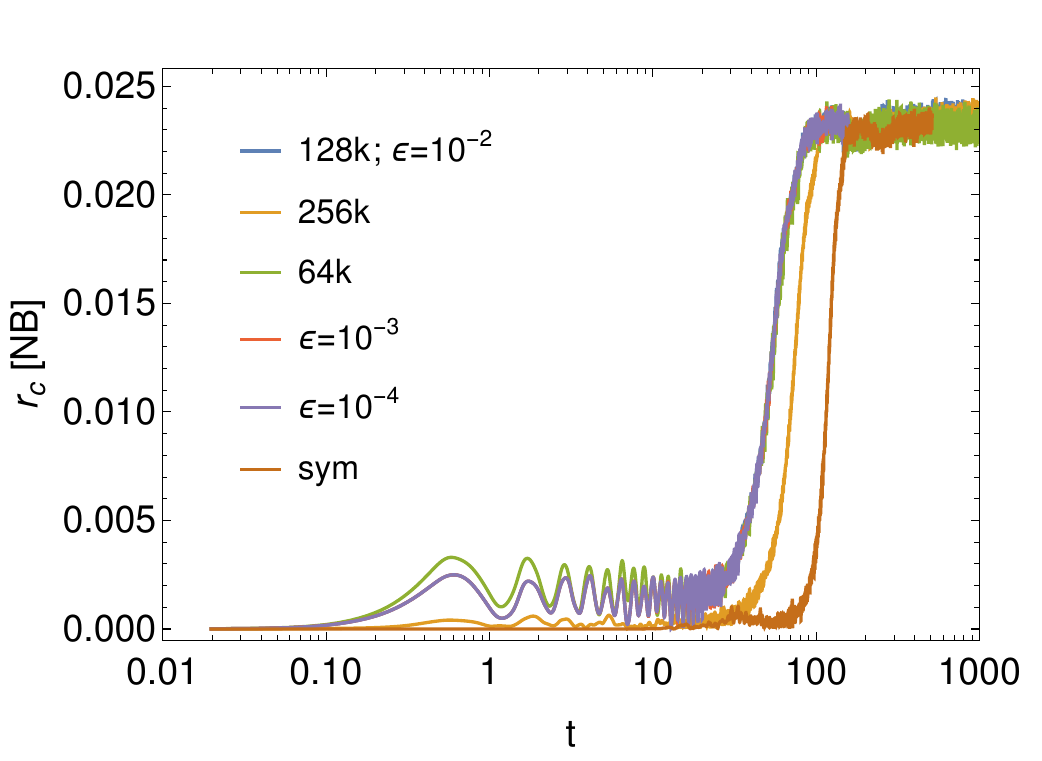}
\caption{Evolution of the radius of the  central mass (logarithmic scale) for run-0.5-60-0.3 for different numerical parameters: number of particles (64k, 128k, 256k), softening parameter ($10^{-2}$–$10^{-4}$), and initial symmetry.}
\label{fig:app_1_1}
\end{figure}

We first varied the number of particles. As shown in Fig.~\ref{fig:app_1_1}, a larger number of particles ($N = 256\mathrm{k}$) leads to smaller fluctuations during the early stages of the evolution (the first tens of orbital periods), due to a smoother gravitational potential. However, once the torus reaches a quasi-equilibrium state (Sect.~\ref{subsec:virial}), all curves for $r_c$ begin to grow and eventually converge to the same value.
In the second set of runs, we varied the softening parameter. As seen from Fig.~\ref{fig:app_1_1}, this variation also does not affect the resulting evolution.
Finally, we performed an additional run with an artificially symmetrized initial distribution (“sym”), in which particle positions and velocities were reflected with respect to a chosen azimuthal axis. In this case, the initial evolution shows almost no significant oscillations. Nevertheless, once the torus reaches equilibrium, the radius of the central-mass orbit begins to increase and, in the saturated state, attains the same value as in the other test runs.
We therefore conclude that these variations do not affect the results presented in the main text.  

\section{Orbital-element analysis}
\label{app:2}

Using the orbital-element representation introduced in Sect.~\ref{sec:modes}, we examine the eccentricity distribution and the evolution of individual particle orbits in the nonlinear state.

\subsection{Eccentricity distribution} 
\label{app:ecc_distr}

For runs that reach the saturated regime (see Table~\ref{Table:modes}), the torus is characterized by a nearly universal distribution of orbital eccentricities. Despite substantial differences in the initial conditions, all these runs converge toward nearly the same profile once the torus reaches equilibrium.
\begin{figure}[h!]
\centering
\includegraphics[width = 62mm]{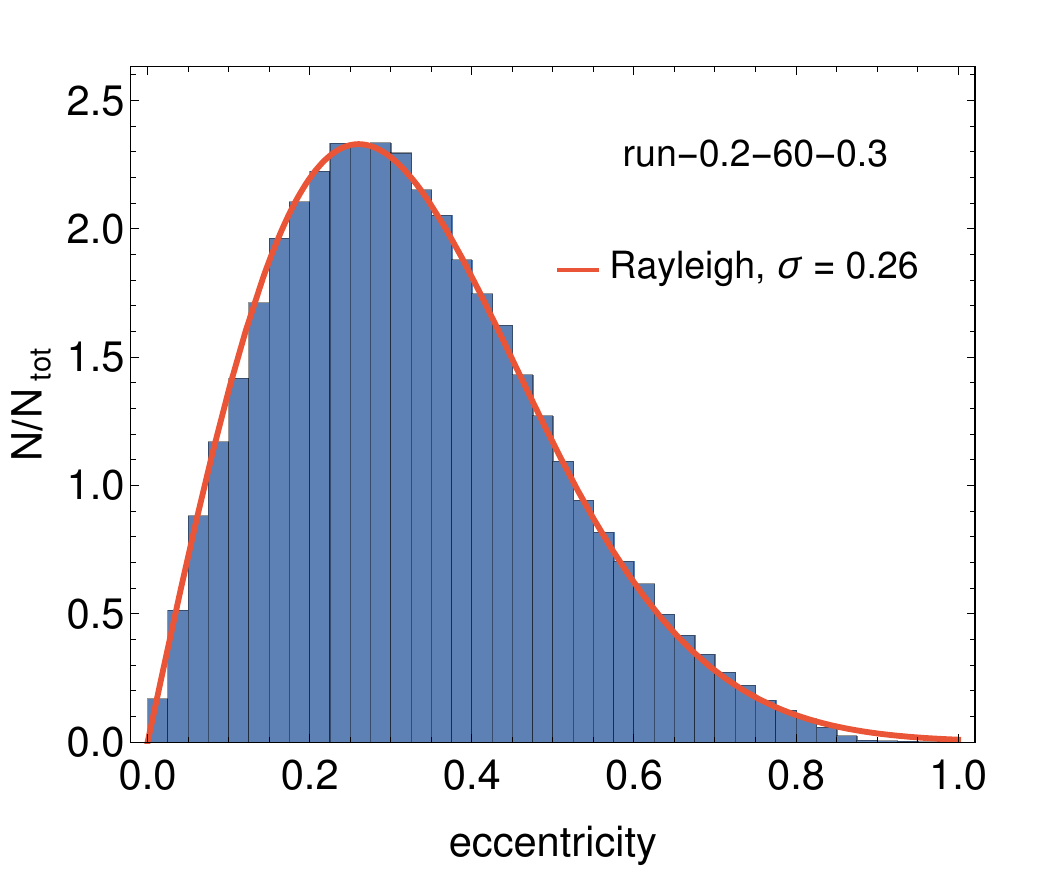}
\includegraphics[width = 62mm]{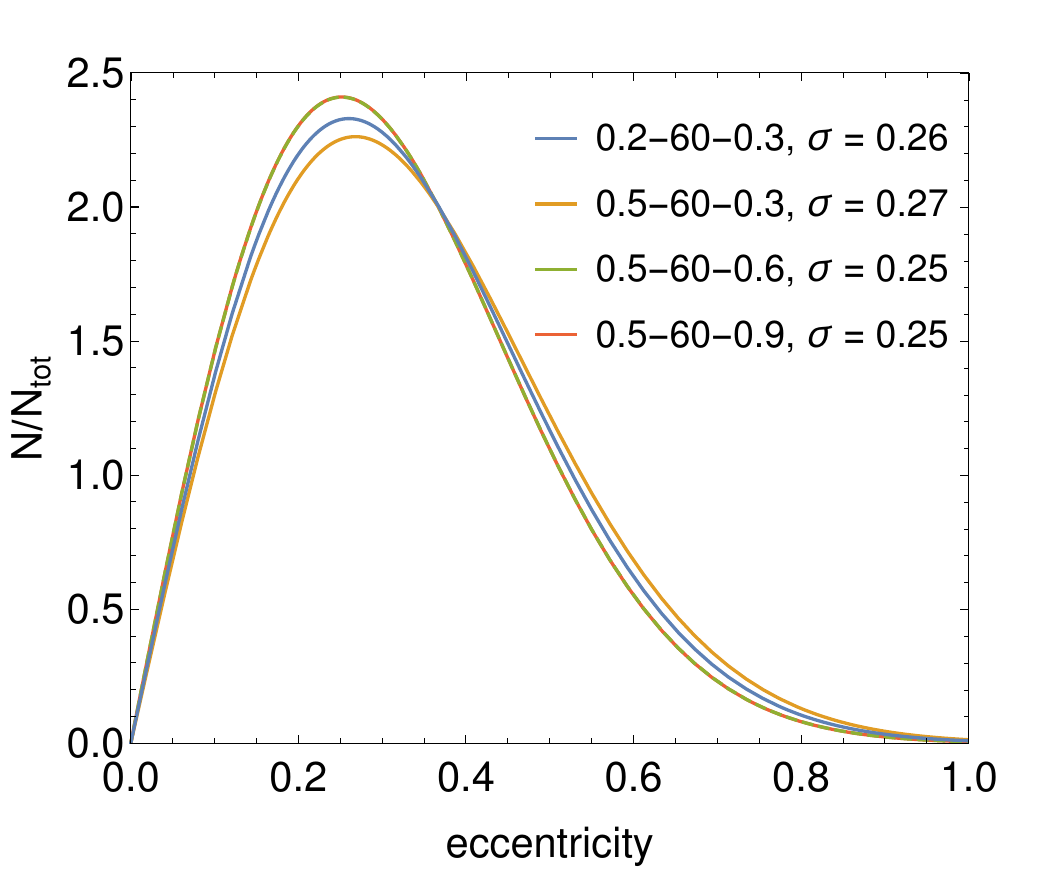}
\caption{The distribution of the particles in eccentricities. Top panel shows the histogram and Rayleigh fit (red curve) for run-0.2-60-0.3 at the 1000 orbital period. Bottom panel shows Rayleigh fits for runs corresponding saturated regime in Table~\ref{Table:modes}.}
\label{fig:app_2_2}
\end{figure}
The final distribution is well described by a Rayleigh law\footnote{Such distributions arise when the components of the eccentricity vector are approximately Gaussian, as commonly found in simulations of gravitationally interacting planetesimal disks \citep{1992Icar...96..107I,2000Icar..143...28S}.},
\begin{equation}
f(e) = \frac{e}{\sigma^2}\exp\left(-\frac{e^2}{2\sigma^2}\right),
\end{equation}
as illustrated for run-0.2-60-0.3 (Fig.~\ref{fig:app_2_2}, top). The eccentricity distributions in the remaining runs are similar and largely overlapping; for clarity, only the corresponding Rayleigh fits are shown (Fig.~\ref{fig:app_2_2}, bottom), with a characteristic eccentricity dispersion $\sigma \approx 0.25$ common to all cases.

This behavior indicates that the system evolves toward a dynamically selected statistical state, in which the torus loses memory of its initial eccentricity distribution and settles into a stable collective configuration. This is consistent with the result of Sect.~\ref{geom_scales}, where all these runs are also characterized by a common scaling relation between harmonic modes.

The stronger overdensity observed in simulations with initially more circular orbits (e.g. $e_{\max}=0.2$) can be understood in terms of the evolution toward the universal equilibrium eccentricity distribution. Although the final distribution is similar in all cases, the required evolution differs depending on the initial conditions. In runs with initially low eccentricities, particles must undergo a larger increase in eccentricity to reach the equilibrium distribution. As a result, a larger fraction of particles is dynamically driven through the high-eccentricity regime associated with the overdensity, allowing more particles to become coherently aligned with the $m=1$ pattern. This leads to a more massive overdensity and a stronger global asymmetry. In contrast, when the initial eccentricity distribution is already broad, fewer particles need to significantly change their eccentricities during the evolution. In this case, a smaller fraction of particles participates in the formation of the overdensity, resulting in a weaker asymmetry, even though the final eccentricity distribution remains similar.

\subsection{Eccentricity--inclination exchange}
\label{subsec:ei}

\begin{figure}[h!]
\centering
\includegraphics[width = 85mm]{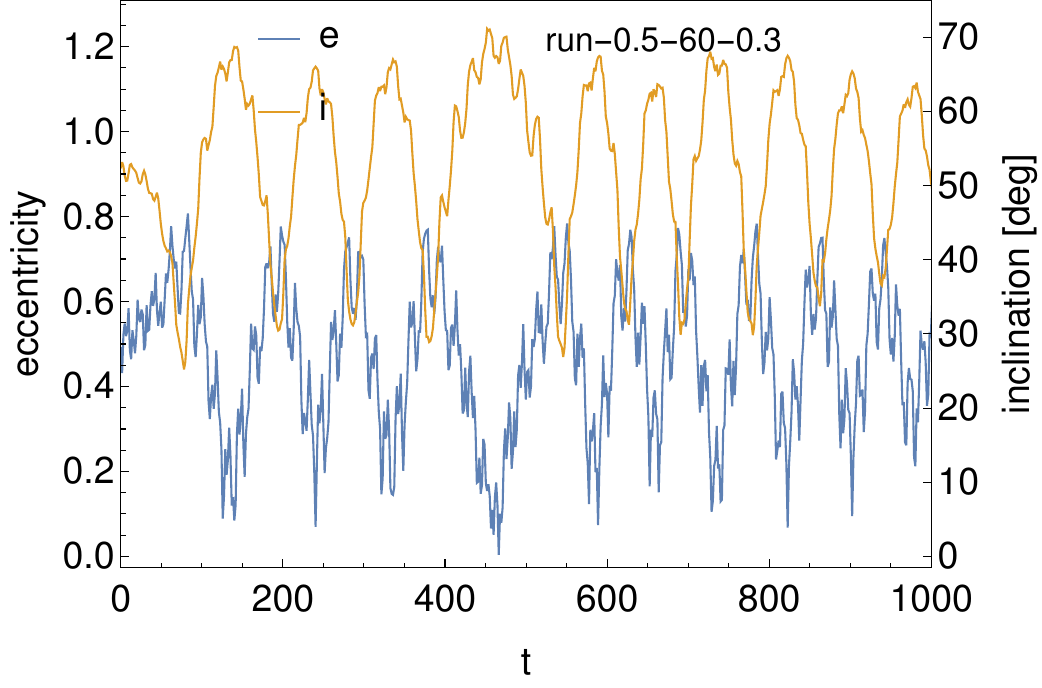}
\caption{Time evolution of the eccentricity and inclination for an individual particle in the torus.}
\label{fig:exc_inc}
\end{figure}

\begin{figure}[h!]
\centering
\includegraphics[width = 85mm]{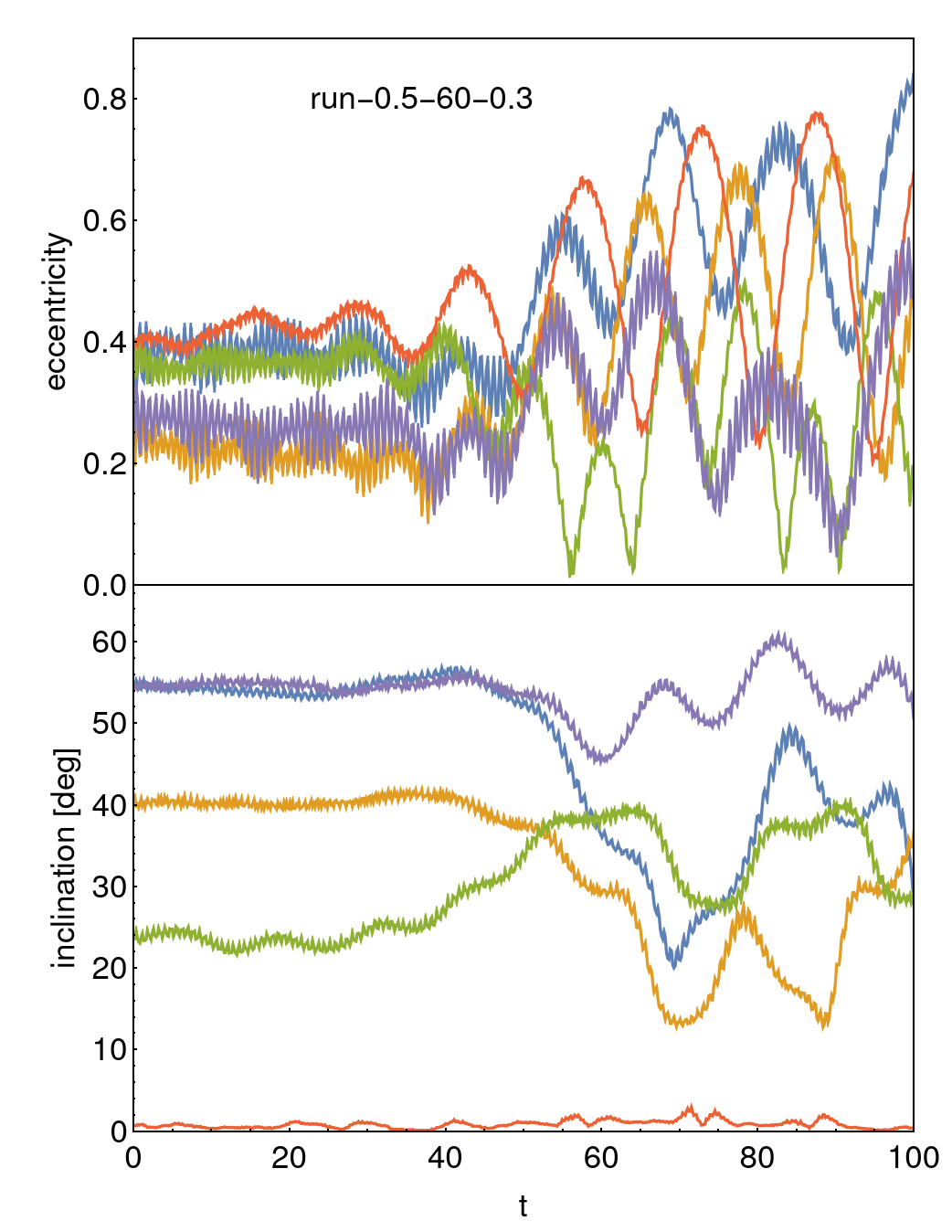}
\caption{Time evolution of the eccentricity and inclination for five particles during the first 100 orbital periods.}
\label{fig:exc_inc_6p}
\end{figure}

To further clarify the dynamical behavior of the system, we examine the time evolution of eccentricities and inclinations of individual particles (for run-0.5-60-0.3). Representative examples are shown in Figs.~\ref{fig:exc_inc} and \ref{fig:exc_inc_6p}. We identify episodes during which an increase in eccentricity is accompanied by a decrease in inclination (Figs.~\ref{fig:exc_inc}), qualitatively similar to Kozai--Lidov oscillations (see \citep{2016ARA&A..54..441N} for a review).

However, the dynamics observed here differs fundamentally from the classical Kozai--Lidov mechanism. In that case, the evolution occurs in a hierarchical three-body system and is governed by a conserved quantity $\sqrt{1-e^2}\cos i$. In our simulations, the gravitational potential is fully self-consistent and time-dependent, shaped by the collective field of the torus and the global $m=1$ pattern. No such invariant is conserved. Moreover, some particles show strong eccentricity growth while their inclination remains nearly constant (Fig.~\ref{fig:exc_inc_6p}), indicating that the evolution is not controlled by a single secular mechanism. The observed behavior reflects the three-dimensional nature of the self-gravitating torus. The non-axisymmetric ($m=1$) gravitational field efficiently modifies the angular momentum of particle orbits, leading to changes in eccentricity. At the same time, the three-dimensional structure of the system allows for variations in orbital inclination through vertical components of the collective gravitational field. As a result, some particles evolve mainly in eccentricity, while others show coupled changes in eccentricity and inclination. This indicates that radial and vertical degrees of freedom are dynamically connected, but not constrained by a single mechanism. This behavior demonstrates that the maintenance of the $m=1$ mode relies on intrinsically three-dimensional orbital dynamics. The vertical degree of freedom provides additional flexibility, helping to maintain apsidal coherence of the global $m=1$ pattern despite differential precession.

This interpretation is consistent with the result presented in Sect.~\ref{sec:thickness}, where geometrically thin configurations fail to develop or sustain a long-lived $m=1$ mode. When the vertical extent of the torus is reduced, the phase space available for such three-dimensional evolution becomes limited, and the nonlinear mechanisms supporting the overdensity are weakened. The individual orbital evolution presented here therefore supports the conclusion that the persistence of the global $m=1$ mode is intrinsically a three-dimensional phenomenon.

\section{The role of torus self-gravity}
\label{app:C}

In this Appendix we examine the role of torus self-gravity by performing an additional simulation of the canonical model run-0.5-60-0.3, but without including mutual gravitational interactions between particles. In this configuration, each particle interacts only with the central mass, while particle--particle forces are neglected.

\begin{figure}[h!]
\centering
\includegraphics[width = 44mm]{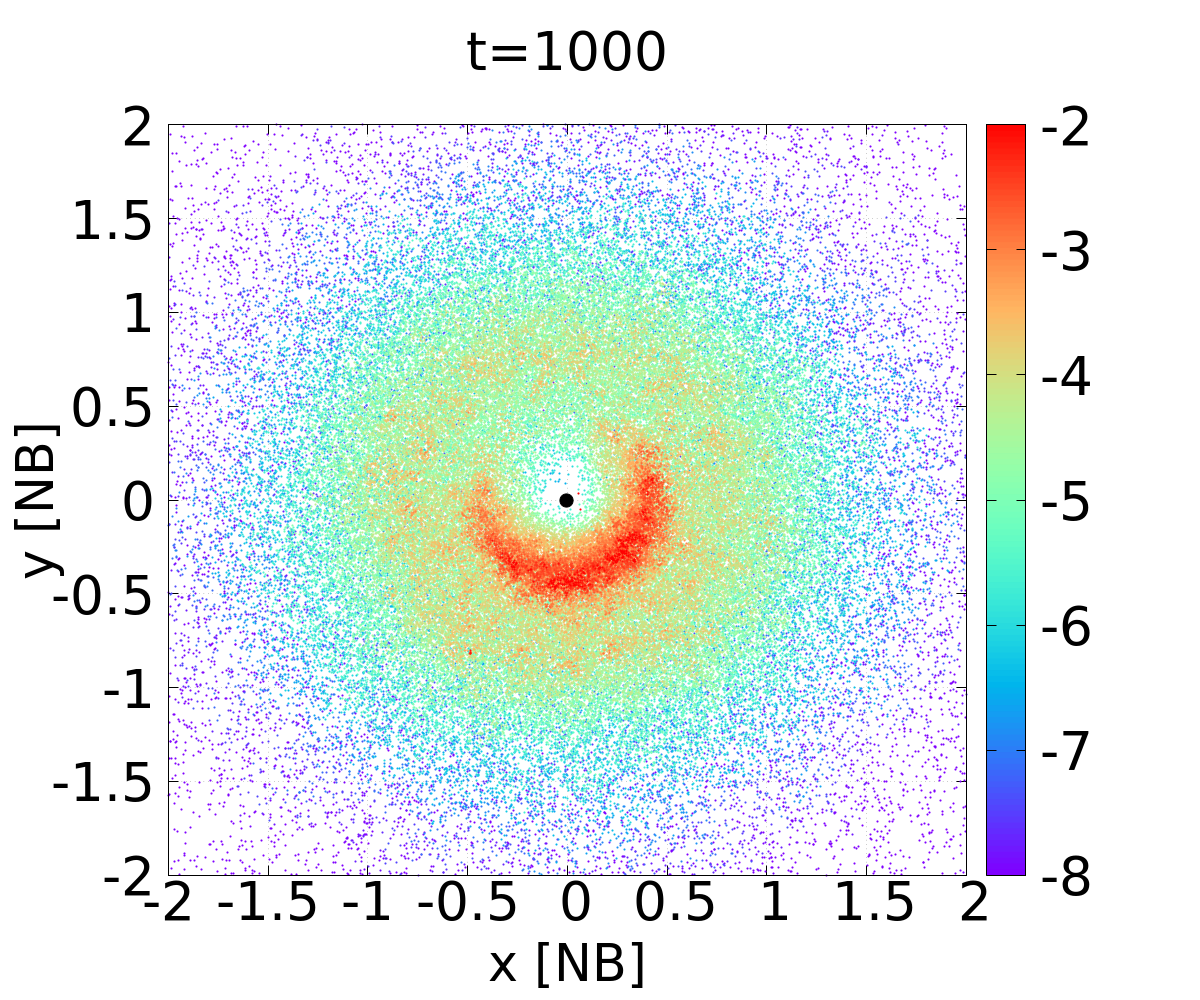}
\includegraphics[width = 44mm]{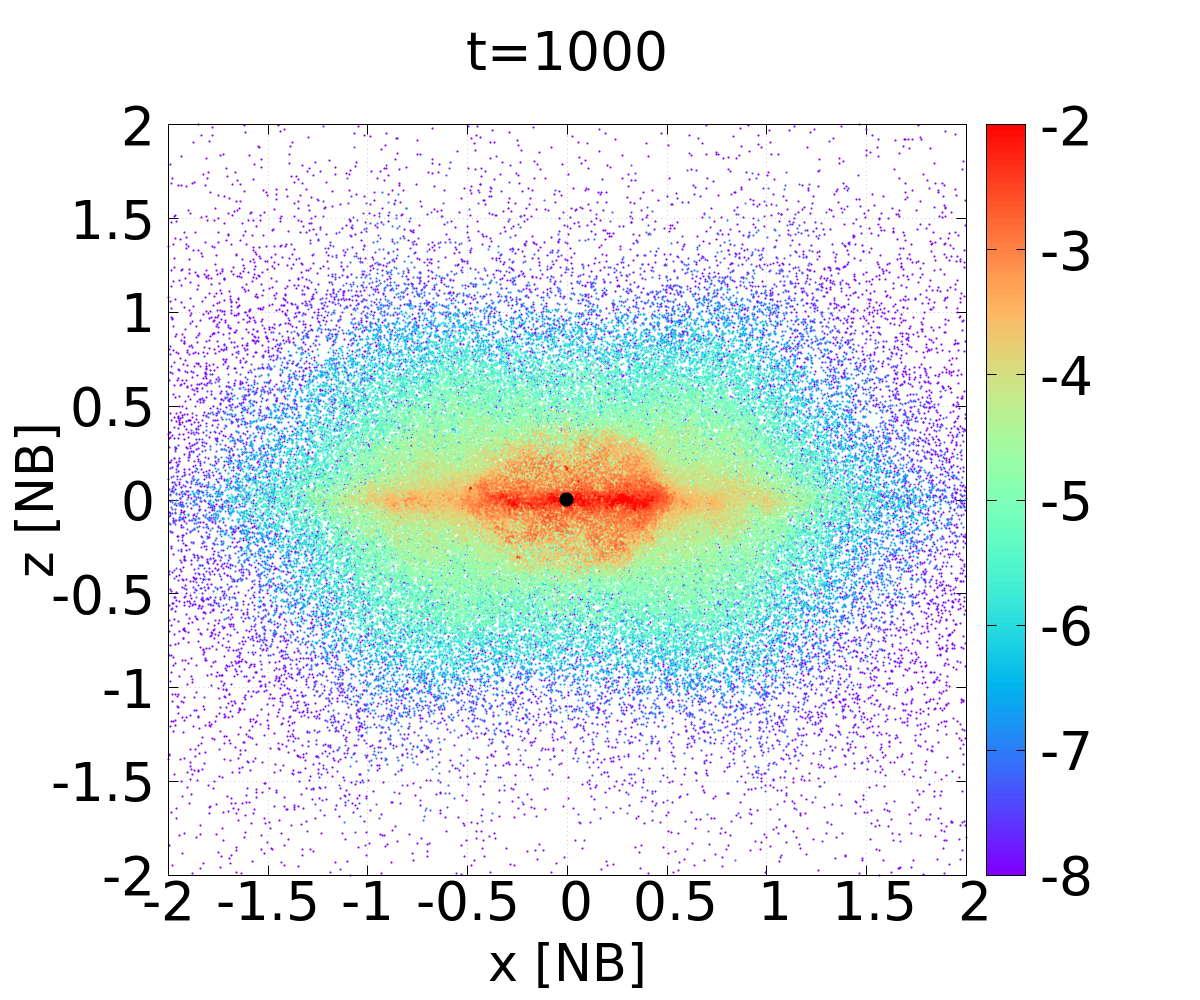}
\caption{The density distribution for the run without taking into account mutual gravitational interaction between particles for the \mbox{$t=1000$}. The initial particle distribution is the same as for run-0.5-60-0.3.}
\label{fig:app_3_1}
\end{figure}
Fig.~\ref{fig:app_3_1} shows the density distribution in this non-self-gravitating case. An apparent overdensity is still visible on the inner side of the torus, taking the form of a diffuse crescent-like feature. However, the particle distribution is significantly more scattered and lacks the sharpness and coherence seen in the fully self-gravitating model. This indicates that the asymmetry is transient and not supported by a self-consistent global mode. The vertical structure (Fig.~\ref{fig:app_3_1}, right panel) also remains diffuse and does not exhibit any coherent large-scale organization. Moreover, in the absence of self-gravity the crescent-like overdensity rotates with a pattern speed comparable to the local orbital frequency. This shows that the structure is not a slow mode, but rather a kinematic feature that follows the motion of individual particles. This behavior contrasts with the self-gravitating case (Fig.~\ref{fig:density_i60}), where a well-defined and long-lived overdensity is sustained. 

\begin{figure}[h!]
\centering
\includegraphics[width = 70mm]{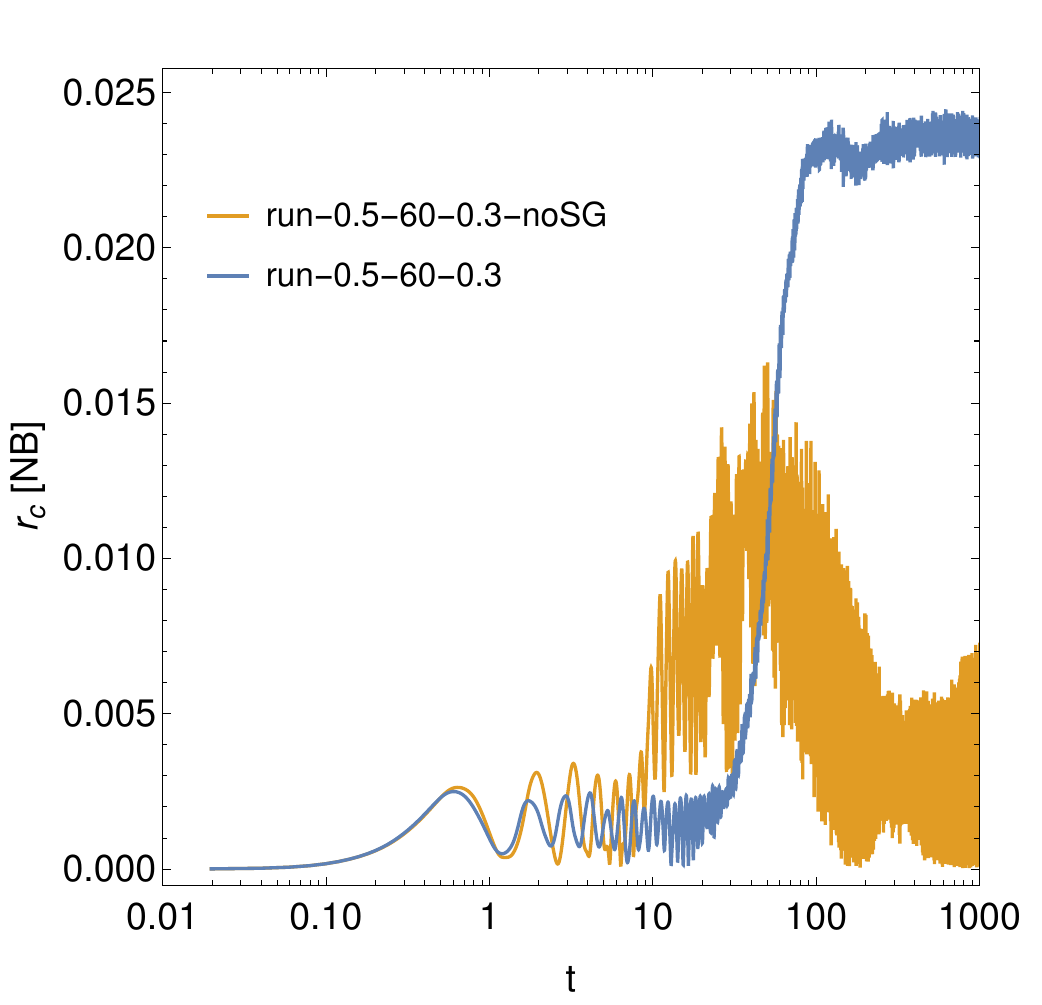}
\caption{The time evolution of the orbital radius of the central mass for two cases. The blue curve corresponds to the case with accounting of self-gravity. The yellow curve corresponds to the case without accounting of self-gravity (noSG).}
\label{fig:app_3_2}
\end{figure}

The dynamical difference between the two cases is further illustrated in Fig.~\ref{fig:app_3_2}, which shows the evolution of the radius of the central mass orbit. Although an $m=1$-like asymmetry initially develops in both cases, the subsequent evolution differs fundamentally. In the non-self-gravitating system (Fig.~\ref{fig:app_3_2}, yellow), the initial growth of the central mass displacement is followed by irregular, strongly fluctuating motion, with no evidence for a stable long-lived configuration. This indicates that the asymmetry can emerge transiently but cannot be sustained, as the system lacks a collective restoring mechanism. In particular, phase coherence between particle orbits cannot be maintained, so differential precession leads to rapid phase mixing and the decay of the overdensity.
We therefore conclude that torus self-gravity is essential for sustaining and regulating a long-lived global $m=1$ mode.

\section{Dependence on the initial virial state}
\label{app:D}

In the canonical model run-0.5-60-0.3, the initial value of the virial
quantity, $2E_{\rm kin}+E_{\rm pot}$, is positive. To verify that the
formation of the global $m=1$ mode is not a consequence of this initial
departure from virial equilibrium, we performed additional simulations
with different initial values of this quantity. In particular, the run
with $e_{\max}=0.3643$ starts close to virial equilibrium, while the runs
with smaller initial eccentricity dispersion, including $e_{\max}=0.1$, start with negative virial value with the amplitude module which is the same as for canonical run.

\begin{figure}[h!]
\centering
\includegraphics[width = 80mm]{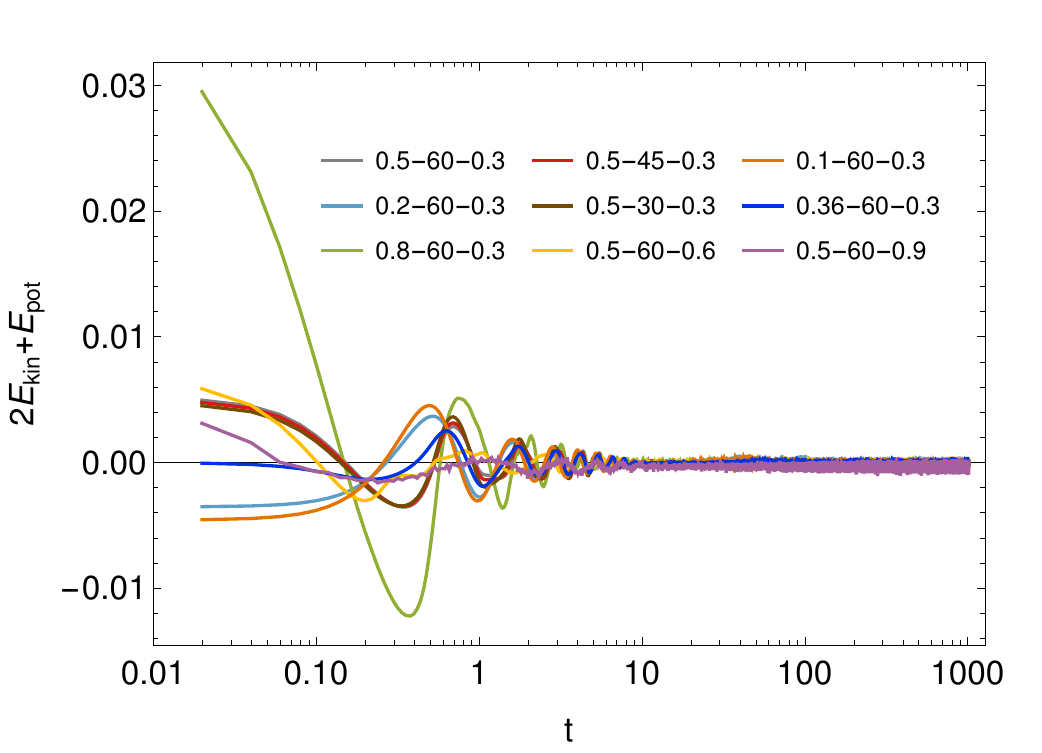}
\caption{The evolution of the virial values for all runs.}
\label{fig:app_4_1}
\end{figure}
Fig.~\ref{fig:app_4_1} shows the evolution of the virial quantity for the full set of runs. Although the initial values differ in both sign and amplitude, all models rapidly relax toward a quasi-virialized state, with subsequent small oscillations around zero. The additional runs with nearly zero and negative initial virial residuals also develop a saturated global $m=1$ mode, as shown by the mode-amplitude evolution in Fig.~\ref{fig:A12-ecc_a}. We therefore conclude that the emergence of the long-lived lopsided mode is insensitive to the initial virial state, which mainly affects the early relaxation stage.

\end{appendix}
\end{document}